\documentclass[11pt,a4paper]{article}
\usepackage{amsmath,amsthm,amsfonts,amssymb,mathrsfs}
\usepackage{graphicx}
\usepackage{subfigure}
\usepackage[latin1]{inputenc}
\usepackage[swedish,english]{babel}
\newcommand{\myvec}[1]{{\boldsymbol#1}}
\allowdisplaybreaks

\begin{document}
\title{Physical-density integral equation methods for scattering
  from multi-dielectric cylinders}
\author{Johan Helsing\thanks{Centre for Mathematical Sciences, Lund
    University, Sweden}~~and Anders Karlsson\thanks{Electrical and
    Information Technology, Lund University, Sweden}}
\date{\today} 
\maketitle

\begin{abstract}
  An integral equation-based numerical method for scattering from
  multi-dielectric cylinders is presented. Electromagnetic fields are
  represented via layer potentials in terms of surface densities with
  physical interpretations. The existence of null-field
  representations then adds superior flexibility to the modeling.
  Local representations are used for fast field evaluation at points
  away from their sources. Partially global representations,
  constructed as to reduce the strength of kernel singularities, are
  used for near-evaluations. A mix of local- and partially global
  representations is also used to derive the system of integral
  equations from which the physical densities are solved. Unique
  solvability is proven for the special case of scattering from a
  homogeneous cylinder under rather general conditions. High
  achievable accuracy is demonstrated for several examples found in
  the literature.
\end{abstract}

\section{Introduction}

Integral equation methods based on local and global integral
representations of electromagnetic fields are presented for the
two-dimensional transmission setting of an incident time harmonic
transverse magnetic wave that is scattered from an object consisting
of an arbitrary number of homogeneous dielectric regions. Several
numerical difficulties are encountered in the evaluations of the
electric and magnetic fields outside and inside the object. That
places high demands on the choice of integral equations, integral
representations, and numerical techniques.

It was seen in~\cite{HelsKarl18}, where scattering from a homogeneous
object was treated, that uniqueness- and numerical problems may occur
for objects having complex permittivities. In that paper the key to
these problems was a system of integral equations for surface
densities without physical interpretation (an indirect formulation
with abstract densities). We now show that uniqueness statements and
accurate field evaluations can also be obtained using a system of
integral equations for surface densities with physical interpretation
(a direct formulation with physical densities). This is an important
step since direct formulations can deliver even higher field accuracy
than can indirect formulations.

Other important results of this paper concern objects that consist of
more than one dielectric region. Three major numerical challenges are
encountered. The first is to accurately evaluate the electric and
magnetic fields close to boundaries. The second is to find and solve
integral equations for objects with boundary triple junctions, and to
evaluate fields close to such points. The third is to accurately
evaluate the electric field when contrasts between regions are very
high. The integral representations and equations we have developed to
meet these challenges are based upon physical surface densities and
global layer potentials. The advantage with our global layer
potentials is that they can be combined to have weaker singularities
in their kernels in more situations than can other layer potentials.

The numerical challenges that remain after our careful modeling are
taken care of by Nyström discretization, accelerated with recursively
compressed inverse preconditioning, and product integration. Numerical
examples constitute an important part of the paper since they verify
that our choices of integral representations and equations are indeed
efficient and can handle all of the difficulties described above.

The present work can be viewed as a continuation of the
work~\cite{HelsKarl18}, on scattering from homogeneous objects, which
uses several results from~\cite{KleiMart88} and~\cite{KresRoac78}. Two
new integral equation formulations have recently been applied to
problems that are similar to the present scattering problem. The first
is referred to as the multi-trace formulation (MTF). It is based upon
a system of Fredholm first-kind equations, that by a Calderón diagonal
preconditioner can be transformed into a system of second-kind
equations~\cite{ClaeHipt13,JHPAT17}. The other is the single-trace
formulation (STF). It is based on a system of Fredholm second-kind
equations for abstract layer potentials~\cite{JHPAT17}. An approach
similar to the STF is presented in~\cite{GreeLee12}. Special attention
is given to numerical problems that arise at triple junctions, and in
that respect Refs.~\cite{ClaeHipt13,GreeLee12,JHPAT17} have much in
common with the present work. A difference is that our representations
lead to cancellation of kernel singularities at field points close to
boundaries, which is important for accurate field evaluations.

The paper is organized as follows: Section~\ref{sec:problem} details
the problem to be solved. Physical surface densities and regional and
global layer potentials and integral operators are reviewed in
Section~\ref{sec:potentials}. Section~\ref{sec:intrep} introduces
local integral representations and null-fields. These are assembled
and used for the construction of integral equations with global
integral operators in Section~\ref{sec:inteq}, which also contains the
proof of unique solvability for the special case of a homogeneous
object. Sections~\ref{sec:eval} and~\ref{sec:evalextend} are on the
evaluation of electromagnetic fields. Section~\ref{sec:sepdist} shows
that certain contributions to global integral representations and
operators are superfluous and can be removed for better numerical
performance. Section~\ref{sec:disc} reviews discretization techniques.
Section~\ref{sec:compare} puts our integral equations into a broader
context by comparing them with popular formulations for the Maxwell
transmission problem in three dimensions.  Our methods are then tested
in three well-documented numerical examples in
Section~\ref{sec:numex}. Section~\ref{sec:conclus} contains
conclusions.

\section{Problem formulation}
\label{sec:problem}

This section presents the problem we shall solve as a partial
differential equation (PDE) and reviews relations between magnetic and
electric fields.

\subsection{Geometry and unit vectors}

The geometry is in $\mathbb{R}^2$ and consists of a bounded object
composed of $N-1$ dielectric regions $\Omega_n$, $n=2,\ldots,N$, which
is surrounded by an unbounded dielectric region $\Omega_1$. A point in
$\mathbb{R}^2$ is denoted $r=(x,y)$. Each region $\Omega_n$ has unit
relative permeability and is characterized by its relative
permittivity $\varepsilon(r)=\varepsilon_n$, $r\in\Omega_n$,
$n=1,\ldots,N$.

\begin{figure}[t]
\centering
\includegraphics[height=50mm]{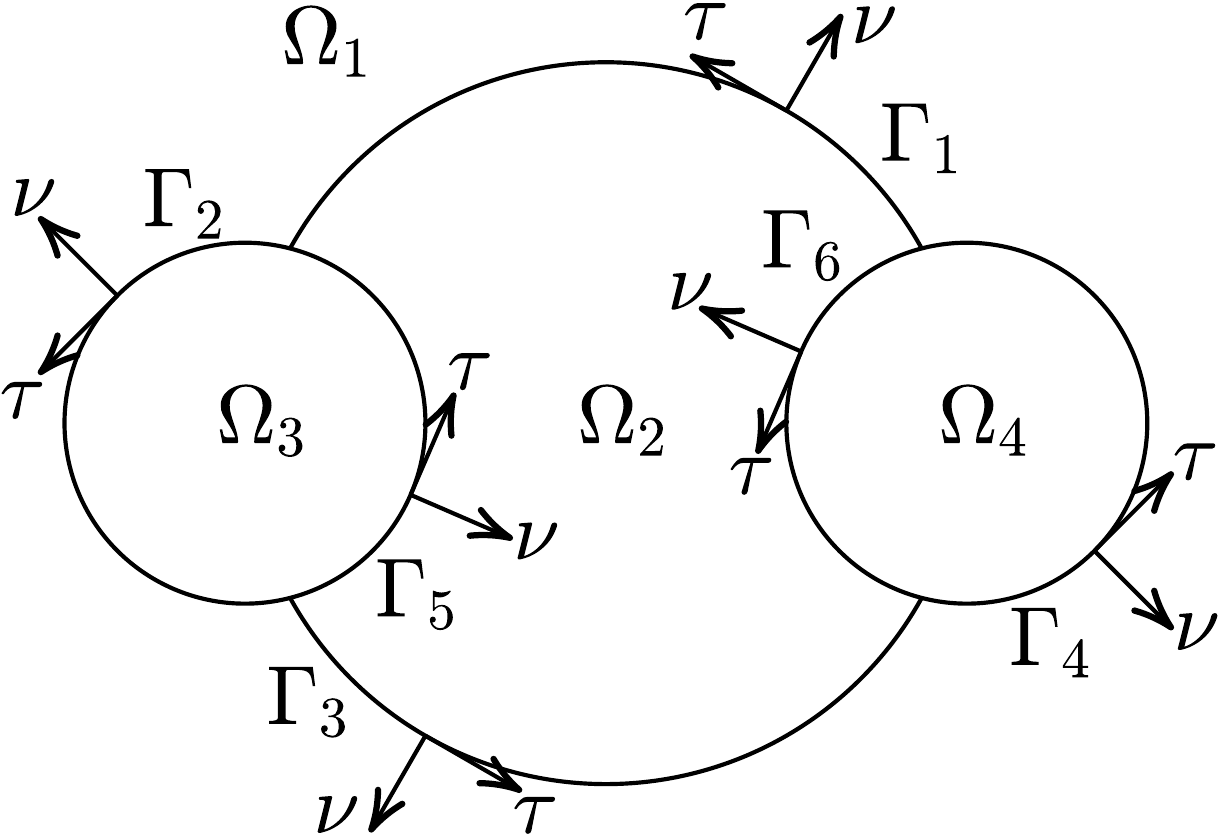}
\caption{\sf The boundaries $\Gamma$ of a four-region configuration with 
  tangential and normal unit vectors $\tau$ and $\nu$. The closed
  curves ${\cal C}_n$ consist of subcurves $\Gamma_m$: ${\cal
    C}_1=\Gamma_1\cup\Gamma_2\cup\Gamma_3\cup\Gamma_4$, ${\cal
    C}_2=\Gamma_1\cup\Gamma_5\cup\Gamma_3\cup\Gamma_6$, ${\cal
    C}_3=\Gamma_2\cup\Gamma_5$, and ${\cal
    C}_4=\Gamma_4\cup\Gamma_6$.}
\label{fig:amoeba2}
\end{figure}

The regions $\Omega_n$, $n\geq 2$, are bounded by closed curves ${\cal
  C}_n$ that consist of subcurves $\Gamma_m$, see
Figure~\ref{fig:amoeba2} for an example. The total number of subcurves
is $M$ and their union is denoted $\Gamma$. The closed curve ${\cal
  C}_1$, bounding the region $\Omega_1$, is the outer boundary of the
object (the point at infinity is not included). If a closed curve
${\cal C}_n$ is traversed so that $\Omega_n$ is on the right, we say
that ${\cal C}_n$ is traversed in a {\it clockwise } direction. The
opposite direction is called {\it counterclockwise}. Note that this
definition of clockwise and counterclockwise agrees with intuition for
all ${\cal C}_n$, seen as isolated closed curves, except for ${\cal
  C}_1$.

Each subcurve $\Gamma_m$ has a tangential unit vector
$\tau=(\tau_x,\tau_y)$ that defines the orientation of $\Gamma_m$ and
a normal unit vector $\nu=(\nu_x,\nu_y)$ that points to the right with
respect to the orientation of $\Gamma_m$, as in Figure
\ref{fig:amoeba2}. We shall also use the standard basis in
$\mathbb{R}^3$ with $\hat{\myvec x}=(1,0,0)$, $\hat{\myvec
  y}=(0,1,0)$, and $\hat{\myvec z}=(0,0,1)$. The vectors $\tau$ and
$\nu$ are related by
\begin{equation}
\myvec\tau=\hat{\myvec z}\times\myvec\nu,
\end{equation}
where $\myvec\tau=(\tau_x,\tau_y,0)$ and $\myvec\nu=(\nu_x,\nu_y,0)$.

\subsection{PDE formulation of the transmission problem}
\label{sec:PDE}

The aim is to find the magnetic and electric fields, $\myvec H(r)$ and
$\myvec E(r)$, in all regions $\Omega_n$, given an incident
time-harmonic transverse magnetic (TM) plane wave. Both $\myvec E$ and
$\myvec H$ can be expressed in terms of a scalar field $U(r)$. The
vacuum wavenumber is denoted $k_0$ and the wavenumbers in the regions
are
\begin{align}
k(r)=k_n\equiv\sqrt{\varepsilon_n}k_0\,,\quad r\in\Omega_n\,,
\quad n=1,\ldots,N\,.
\label{eq:k2k1}
\end{align}
For $r\in\Gamma_m$ we define $U^{\rm R}(r)$, $\nabla U^{\rm R}(r)$,
$k_m^{\rm R}$, and $\varepsilon_m^{R}$ as the limit scalar field, the
limit gradient field, the wavenumber, and the permittivity on the
right-hand side of $\Gamma_m$. On the left-hand side of $\Gamma_m$ the
corresponding quantities are $U^{\rm L}(r)$, $\nabla U^{\rm L}(r)$,
$k_m^{\rm L}$, and $\varepsilon_m^{L}$.

The PDE for the scalar field is
\begin{align}
\Delta U(r)+k(r)^2 U(r)&=0\,,\quad r\in\mathbb{R}^2\setminus\Gamma\,,
\label{eq:Hlm}
\end{align}
with boundary conditions on $\Gamma_m$, $m=1,\ldots,M$,
\begin{align}
U^{\rm R}(r)&=U^{\rm L}(r)\,,\quad r\in\Gamma_m\,,
\label{eq:BC1}\\
(\varepsilon_m^{\rm R})^{-1}\nu\cdot\nabla U^{\rm R}(r)&=(\varepsilon_m^{\rm L})^{-1}\nu\cdot\nabla U^{\rm L}(r)\,,\quad r\in\Gamma_m\,.
\label{eq:BC2}
\end{align}
In $\Omega_1$ the field is decomposed into an incident and a scattered field
\begin{equation}
  U(r)=U^{\rm in}(r)+U^{\rm sc}(r)\,,\quad r\in\Omega_1,
  \end{equation}
where
\begin{align}
\Delta U^{\rm in}(r)+k_1^2 U^{\rm in}(r)&=0\,,\quad r\in \mathbb{R}^2\,.
\label{eq:Incident}
\end{align}
The scattered field satisfies the radiation condition  
\begin{equation}
 \lim_{|r|\to\infty}\sqrt{|r|}
\left(\frac{\partial}{\partial|r|}-{\rm i}k_1\right)U^{\rm sc}(r)=0\,,
\quad r\in\Omega_1\,.
\label{eq:Hlm6}
\end{equation}
The time dependence $e^{-{\rm i}t}$ is assumed and the angular
frequency is scaled to one.

\subsection{The magnetic and electric fields}

The complex magnetic and electric fields are
\begin{align}
\myvec H(r)&=U(r)\hat{\myvec z}\,,\quad r\in\mathbb{R}^2\,,
\label{eq:H}\\
\myvec E(r)&={\rm i}k_0^{-1}\varepsilon_n^{-1}\nabla_3 U(r)
\times\hat{\myvec z}\,,\quad r\in\Omega_n\,,
\label{eq:E}
\end{align}
where the $\nabla_3 U(r)$ is the gradient $\nabla U(r)$ extended with
a zero third component. The corresponding time-domain fields are
\begin{align}
\myvec H(r,t)&=\Re{\rm e}\{\myvec H(r)e^{-{\rm i}t}\}\,,\\
\myvec E(r,t)&=\Re{\rm e}\{\myvec E(r)e^{-{\rm i}t}\}\,.
\label{eq:HandEtime}
\end{align}
The electric field is scaled with the wave impedance of vacuum,
$\eta_0$, in order to give the magnetic and electric fields the same
dimension.

\subsection{The incident plane wave}

The incident TM-wave travels in the direction $d=(d_x,d_y)$, and the
scalar field $U^{\rm in}$, magnetic field $\myvec H^{\rm in}$, and
electric field $\myvec E^{\rm in}$ of this wave are
\begin{align}
 &U^{\rm in}(r)=e^{{\rm i}k_1(d\cdot r)}\,,\\
 &\myvec H^{\rm in}(r)=e^{{\rm i}k_1(d\cdot r)}\hat{\myvec z}\,,\\
 &\myvec E^{\rm in}(r)=
     \dfrac{1}{\sqrt{\varepsilon_1}}(-d_y,d_x,0)e^{{\rm i}k_1(d\cdot r)}\,.
\end{align}
The normal component of the incident electric field on a boundary is
\begin{equation}
\myvec\nu\cdot\myvec E ^{\rm in}(r)=
\dfrac{1}{\sqrt{\varepsilon_1}}(\nu_yd_x-\nu_xd_y)e^{{\rm i}k_1(d\cdot r)}\,.
\end{equation}

\section{Physical densities, potentials, and operators}
\label{sec:potentials}

Two surface densities, $\mu(r)$ and $\rho(r)$, are introduced and
referred to as {\it physical densities}. The density $\mu$ is the
tangential component of $\myvec H$ and the density $\rho$ is
proportional to the tangential component of $\myvec E$
\begin{gather}
\mu(r) =U(r)\,,
\quad r\in\Gamma_m\,,
\label{eq:mu}\\
\rho(r)=(\varepsilon_m^{\rm R})^{-1}\nu\cdot\nabla U^{\rm R}(r)\,,
\quad r\in\Gamma_m\,.
\label{eq:rhoR}
\end{gather}
If $U$ is a solution to the transmission problem of
Section~\ref{sec:PDE} it also holds, because of~(\ref{eq:BC2}), that
\begin{equation}
\rho(r)=(\varepsilon_m^{\rm L})^{-1}\nu\cdot\nabla U^{\rm L}(r)\,,
\quad r\in\Gamma_m\,.
\label{eq:rhoL}
\end{equation}

In~\cite{HelsKarl18} it was shown that integral equation-based
numerical schemes involving physical densities offer certain
advantages for field evaluations close to $\Gamma$, compared to
indirect schemes involving surface densities without immediate
physical interpretations ({\it abstract densities}). We here pursue
the concept of physical densities.

\subsection{Acoustic layer potentials and operators}

The fundamental solution to the Helmholtz equation is taken as
\begin{equation}
\Phi_k(r,r')=\frac{\rm i}{2}H_0^{(1)}(k|r-r'|)\,,
\label{eq:green}
\end{equation}
where $H_0^{(1)}$ is the zeroth order Hankel function of the first
kind. On each subcurve $\Gamma_m$ we need six right-hand acoustic
layer potentials defined in terms of a general surface density
$\sigma(r)$ as
\begin{align}
S_m^{\rm R}\sigma(r)&=
 \int_{\Gamma_m}\Phi_{k_m^{\rm R}}(r,r')\sigma(r')\,{\rm d}\ell'\,,\\
K_m^{\rm R}\sigma(r)&=
 \int_{\Gamma_m}\frac{\partial\Phi_{k_m^{\rm R}}}{\partial\nu'}(r,r')\sigma(r')
  \,{\rm d}\ell'\,,\\
K^{\rm RA}_m\sigma(r)&=
 \int_{\Gamma_m}\frac{\partial\Phi_{k_m^{\rm R}}}{\partial\nu}(r,r')\sigma(r')
  \,{\rm d}\ell'\,,\\
T_m^{\rm R}\sigma(r)&=
  \int_{\Gamma_m}\frac{\partial^2\Phi_{k_m^{\rm R}}}
  {\partial\nu\partial\nu'}(r,r')\sigma(r')\,{\rm d}\ell',\\
\myvec B_m^{\rm R}\sigma(r)&=
  \int_{\Gamma_m}\Phi_{k_m^{\rm R}}(r,r')
  \myvec\tau(r')\sigma(r')\,{\rm d}\ell'\,,\\
C_m^{\rm R}\sigma(r)&=
  \int_{\Gamma_m}\frac{\partial\Phi_{k_m^{\rm R}}}
  {\partial\tau}(r,r')\sigma(r')\,{\rm d}\ell'\,.
\end{align}

Here $r\in \mathbb{R}^2$, $\partial/\partial\nu'=\nu(r')\cdot\nabla'$,
$\partial/\partial\nu=\nu(r)\cdot\nabla$,
$\partial/\partial\tau=\tau(r)\cdot\nabla$, and we have extended the
definition of the rightward unit normal $\nu=\nu(r)$ at a point
$r\in\Gamma$ so that if $r\notin\Gamma$, then $\nu$ is to be
interpreted as an arbitrary unit vector associated with $r$. The
left-hand layer potentials $S_m^{\rm L}\sigma(r)$, $K_m^{\rm
  L}\sigma(r)$, $K_m^{\rm LA}\sigma(r)$, $T_m^{\rm L}\sigma(r)$,
$\myvec B_m^{\rm L}\sigma(r)$, and $C_m^{\rm L}\sigma(r)$ are defined
analogously to the right-hand potentials.

For each closed curve ${\cal C}_n$ we now define the six regional
layer potentials $S_n\sigma(r)$, $K_n\sigma(r)$, $K^{\rm
  A}_n\sigma(r)$, $T_n\sigma(r)$, $\myvec B_n\sigma(r)$, and
$C_n\sigma(r)$ via
\begin{equation}
G_n\sigma(r)=
 \sum\limits_{{\rm ccw}\,\Gamma_m\in{\cal C}_n}G_m^{\rm L}\sigma(r)  
-\sum\limits_{{\rm cw}\,\Gamma_m\in{\cal C}_n}G_m^{\rm R}\sigma(r)\,, 
\quad r\in\mathbb{R}^2\,,
\label{eq:Gopern}
\end{equation}
where $G$ can represent $S$, $K$, $K^{\rm A}$, $T$, $\myvec B$, and
$C$ and where ``${\rm cw}\,\Gamma_m$'' and ``${\rm ccw}\,\Gamma_m$''
denote subcurves with clockwise and counterclockwise orientations
along ${\cal C}_n$. For the special case of $r\in\Gamma$ we refer to
the $G_n$ of~(\ref{eq:Gopern}) as integral operators.

When $r^{\circ}\in{\cal C}_n$, and with some abuse of notation, one
can show the limits
\begin{equation}
\begin{split}
\lim_{\Omega_n\ni r\to r^\circ} S_n\sigma(r)&=S_n\sigma(r^\circ)\,, \\
\lim_{\Omega_n\ni r\to r^\circ} K_n\sigma(r)&=-\sigma(r^\circ)
                                         +K_n\sigma(r^\circ)\,, \\
\lim_{\Omega_n\ni r\to r^\circ} K_n^{\rm A}\sigma(r)&=\sigma(r^\circ)
                                                  +K_n^{\rm A}\sigma(r^\circ)\,, \\
\lim_{\Omega_n\ni r\to r^\circ} T_n\sigma(r)&=T_n\sigma(r^\circ)\,,\\
\lim_{\Omega_n\ni r\to r^\circ} C_n\sigma(r)&=C_n\sigma(r^\circ)\,,\\
\lim_{\Omega_n\ni r\to r^\circ} \myvec B_n\sigma(r)&=\myvec B_n\sigma(r^\circ)\,.
\end{split}
\label{eq:limits}
\end{equation}
Here $C_n\sigma(r^\circ)$ is to be understood in the Cauchy
principal-value sense and $T_n\sigma(r^\circ)$ in the Hadamard
finite-part sense. See~\cite[Theorem 3.1]{ColtKres98}
and~\cite[Theorem 2.21]{ColtKres83} for more precise statements on
these limits and~\cite[Theorem 5.46]{KirsHett15} for statements in a
modern function-space setting.

\subsection{The singular nature of kernels}
\label{sec:singnat}

In a similar way as in the indirect approach
of~\cite[Section~3]{GreeLee12} and \cite[Section~3.2]{JHPAT17}, we
plan to derive a {\it global integral representation} of the scalar
field $U$. A global representation means that the densities $\mu(r)$
and $\rho(r)$ are used to represent $U$ in every region $\Omega_n$,
whether $r$ is on the boundary of that region or not~\cite{GreeLee12}.
For this, we need to introduce {\it global layer potentials and
  integral operators}, which are sums over their regional
counterparts. This section, which draws
on~\cite[Section~4]{HelsKarl18}, collects known results on the
singular nature of kernels of various potentials and operators that
occur in our representations of $U$ and $\myvec E$ and in our integral
equations for $\mu$ and $\rho$.

The kernels of the global layer potentials
\begin{equation}
\sum_{n=1}^N\varepsilon_nS_n\sigma(r)\,, \quad 
\sum_{n=1}^NK_n\sigma(r)\,,\quad r\in\mathbb{R}^2\setminus\Gamma\,,
\end{equation}
exhibit logarithmic singularities as $r\to r'\in\Gamma$. This is so
since $\Phi_k(r,r')$ of~(\ref{eq:green}) has a logarithmic singularity
as $r\to r'$ and this singularity carries over to the kernel of
$S_n\sigma(r)$. From~(\ref{eq:Gopern}) it follows that
\begin{equation}
\sum_{n=1}^NK_n\sigma(r)=
\sum_{m=1}^M\left (K_m^{\rm L}-K_m^{\rm R}\right)\sigma(r)\,.
\label{eq:Koper2}
\end{equation}
In each term in the sum on the right-hand side of~(\ref{eq:Koper2})
the leading Cauchy-singular parts of the kernels of $K_m^{\rm
  L}\sigma(r)$ and $K_m^{\rm R}\sigma(r)$ are independent of the
wavenumber, see~\cite[Section~4.3]{HelsKarl18}, and cancel out.
Changing the order of summation in global layer potentials, as
in~(\ref{eq:Koper2}), is helpful when studying their singularities.

The kernel of
\begin{equation}
\sum_{n=1}^N\varepsilon_n\partial_iS_n\sigma(r)\,,
\quad r\in\mathbb{R}^2\setminus\Gamma\,,
\quad i=x,y\,,
\end{equation}
exhibits logarithmic- and Cauchy-type singularities as $r\to
r'\in\Gamma$. The kernel of
\begin{equation}
\sum_{n=1}^N\partial_i K_n\sigma(r)\,, 
\quad r\in\mathbb{R}^2\setminus\Gamma\,,
\quad i=x,y\,,
\end{equation}
exhibits, strictly speaking, only logarithmic singularities as $r\to
r'\in\Gamma$. In the context of numerical product integration,
however, it is advantageous to consider this kernel as having a
Cauchy-type singularity. See~\cite[Section 4.5]{HelsKarl18}. The
kernels of $\partial_iS_n\sigma(r)$, $i=x,y\,$, exhibit logarithmic-
and Cauchy-type singularities as $\Omega_n\ni r\to r'\in{\cal C}_n$.
The kernel of $\myvec B_n\sigma(r)$ exhibits logarithmic singularities
as $\Omega_n\ni r\to r'\in{\cal C}_n$.

The global integral operators
\begin{equation}
\sum_{n=1}^NS_n\,, \quad 
\sum_{n=1}^NT_n\,, \quad 
\sum_{n=1}^NC_n\,, \quad
\sum_{n=1}^N\myvec B_n\,,
\label{eq:STCB}
\end{equation}
have weakly singular (logarithmic) kernels and are compact, while
\begin{equation}
\sum_{n=1}^N\varepsilon_n^{-1}K_n\,, \quad
\sum_{n=1}^N\varepsilon_nK_n^{\rm A}\,, \quad
\sum_{n=1}^N\varepsilon_n^{-1}K_n^{\rm A}\,,
\end{equation}
are merely bounded. Away from singular boundary points, such as
corners or triple junctions, these latter operators also have weakly
singular (logarithmic) kernels and are compact. See~\cite[Lemmas
1-2]{LaiJian18} for similar statements on boundaries of simply
connected Lipschitz domains.

\section{Integral representations of $U$ and $\nu\cdot\nabla U$}
\label{sec:intrep}

If $U$ is a solution to the transmission problem of
Section~\ref{sec:PDE}, Green's theorem and~(\ref{eq:mu}),
(\ref{eq:rhoR}), and (\ref{eq:rhoL}), give the {\it local integral
  representation}
\begin{equation}
U(r)=U^{\rm in}(r)\delta_{n1}-\frac{1}{2}\left(K_n\mu(r)
-\varepsilon_nS_n\rho(r)\right)\,,\quad r\in \Omega_n\,,
\label{eq:rep1}
\end{equation}
see~\cite[Section~3.1]{Kris16} and~\cite[Section~4.2]{KleiMart88}. A
local representation means that only the parts of the densities $\mu$
and $\rho$ that are present on ${\cal C}_n$ are used to represent $U$
in $\Omega_n$. For $r$ outside $\Omega_n$ Green's theorem gives
\begin{equation}\begin{split}
0=&U^{\rm in}(r)\delta_{n1}-\frac{1}{2}\left(K_n\mu(r)
-\varepsilon_nS_n\rho(r)\right)\,,\quad r\notin \Omega_n\cup {\cal C}_n\,.
\end{split}
\label{eq:rep2}
\end{equation}
This relation is well-known, but by many different names, such as the
extinction theorem, the Ewald-Oseen extinction
theorem~\cite[Chapter~2.4]{BornWolf65}, the Helmholtz
formulae~\cite{Martin80}, the null-field
equation~\cite{KleiRoacStro84}, and the extended boundary
condition~\cite{Waterman65}. We can add the right-hand side of
\eqref{eq:rep2} to the field in regions outside $\Omega_n$ without
altering the field. This opens up possibilities to weaken
singularities in integral equations and near singularities in integral
representations. In what follows we often use this opportunity to
improve accuracy in the evaluation of magnetic and electric fields.

The directional derivative of \eqref{eq:rep1} and \eqref{eq:rep2}  are
\begin{equation}
\nu\cdot\nabla U(r)=\nu\cdot\nabla U^{\rm in}(r)\delta_{n1}
 -\frac{1}{2}\left(T_n\mu(r)-\varepsilon_nK_n^{\rm A}\rho(r)\right),
\quad r\in\Omega_n\,,
\label{eq:rep3}
\end{equation}
and 
\begin{equation}
0=\nu\cdot\nabla U^{\rm in}(r)\delta_{n1}
-\frac{1}{2}\left(T_n\mu(r)-\varepsilon_nK_n^{\rm A}\rho(r)\right),
\quad r\notin\Omega_n\cup{\cal C}_n\,,
\label{eq:rep4}
\end{equation}
where $\nu=\nu(r)$ is an arbitrary unit vector associated with $r$.

In summary we can say that if $U$ is a solution to the transmission
problem of Section~\ref{sec:PDE}, then~(\ref{eq:rep1}),
(\ref{eq:rep2}), (\ref{eq:rep3}), and (\ref{eq:rep4}) hold for
$r\in\mathbb{R}^2\setminus\Gamma$.

\section{Integral equations}
\label{sec:inteq}

When $\Omega_n\ni r\to {\cal C}_n$ in \eqref{eq:rep1} and
\eqref{eq:rep3}, each boundary ${\cal C}_n$ gives rise to two separate
integral equations
\begin{gather}
\mu(r)+K_n\mu(r)-\varepsilon_nS_n\rho(r)=
2U^{\rm in}(r)\delta_{n1}\,,\quad r\in {\cal C}_n\,,
\label{eq:ie1}\\
\varepsilon_n\rho(r)+T_n\mu(r)-\varepsilon_nK_n^{\rm A}\rho(r)=
2\nu\cdot\nabla U^{\rm in}(r)\delta_{n1}\,,\quad r\in {\cal C}_n\,.
\label{eq:ie2}
\end{gather}
The $2N$ equations \eqref{eq:ie1} and \eqref{eq:ie2} and the
null-field representations~(\ref{eq:rep2}) and~(\ref{eq:rep4}) are now
combined into a single system of integral equations. We first show how
this is done for $N=2$, that is for a homogeneous object, and then
proceed to objects with many regions.

\subsection{Integral equations and uniqueness when $N=2$}
\label{sec:unique}

\begin{figure}[t]
\centering 
\includegraphics[height=38mm]{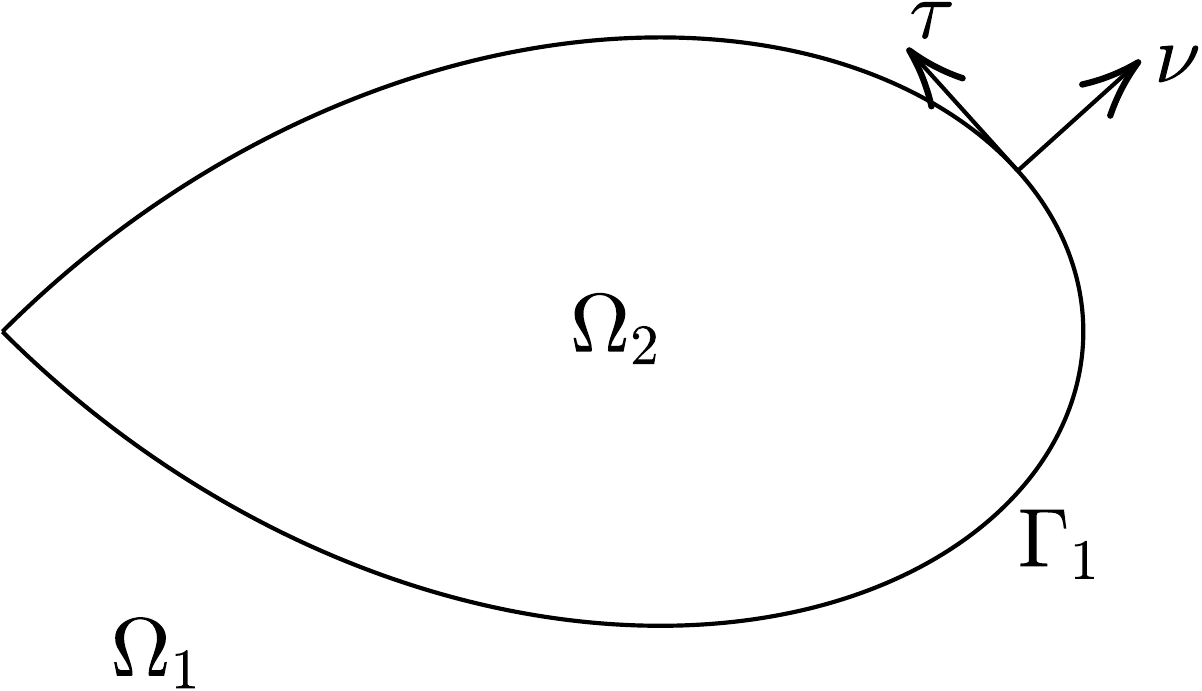}
\caption{\sf The boundary 
  $\Gamma=\Gamma_1={\cal C}_1={\cal C}_2$ of a two-region
  configuration with tangential and normal unit vectors $\tau$ and
  $\nu$.}
\label{fig:amoeba0}
\end{figure}

When $N=2$, then $\Omega_1$ is the outer region, $\Omega_2$ is the
object, and there is only one boundary $\Gamma=\Gamma_1={\cal
  C}_1={\cal C}_2$. See Figure~\ref{fig:amoeba0} for an example. First
we add $\varepsilon_1^{-1}$ times \eqref{eq:ie1} for ${\cal C}_1$ and
$c\varepsilon_2^{-1}$ times \eqref{eq:ie1} for ${\cal C}_2$. Here $c$
is a free parameter such that $c\varepsilon_1+\varepsilon_2\ne0$. This
gives
\begin{equation}
\mu(r)+
 \breve{\alpha}_1\left(\varepsilon_1^{-1} K_1+c\varepsilon_2^{-1}K_2\right)\mu(r)
-\breve{\alpha}_1\left(S_1+cS_2\right)\rho(r)
 =f_1(r)\,,\quad r\in\Gamma\,,
\label{eq:combined1}
\end{equation}
where
\begin{equation}
\breve{\alpha}_1=\frac{\varepsilon_1\varepsilon_2}{c\varepsilon_1+\varepsilon_2}\,,
\quad
f_1(r)=2\breve{\alpha}_1\varepsilon_1^{-1}U^{\rm in}(r)\,.
\end{equation}
The other integral equation is the sum of the ${\cal C}_1$ and ${\cal
  C}_2$ versions of \eqref{eq:ie2}
\begin{equation}
 \rho(r)+\beta_1(T_1+T_2)\mu(r)
-\beta_1\left(\varepsilon_1K_1^{\rm A}+\varepsilon_2K_2^{\rm A}\right)\rho(r)
 =f_2(r)\,,\quad r\in\Gamma\,,
\label{eq:combined2}
\end{equation}
where
\begin{equation}
\beta_1=\frac{1}{\varepsilon_1+\varepsilon_2}\,,
\quad 
f_2(r)=2\beta_1\nu\cdot \nabla U^{\rm in}(r)\,.
\end{equation}

We now write the system of integral equations~(\ref{eq:combined1})
and~(\ref{eq:combined2}) in block-matrix form
\begin{equation}
\begin{bmatrix}
I+\breve{\alpha}_1\left(\varepsilon_1^{-1} K_1+c\varepsilon_2^{-1}K_2\right) 
&-\breve{\alpha}_1\left(S_1+cS_2\right)\\
\beta_1(T_1+T_2)
&I-\beta_1\left(\varepsilon_1K_1^{\rm A}+\varepsilon_2K_2^{\rm A}\right)
\end{bmatrix} 
\begin{bmatrix}
\mu(r)\\
\rho(r)
\end{bmatrix} 
=
\begin{bmatrix}
 f_1(r)\\
 f_2(r)
\end{bmatrix},
\label{eq:HKsys}
\end{equation}
where the diagonal blocks contain global integral operators that are
compact away from singular boundary points and the off-diagonal blocks
contain global operators that are everywhere compact. In particular,
since a clockwise direction for ${\cal C}_1$ is a counterclockwise
direction for ${\cal C}_2$ and vice versa, the operator $T_1+T_2$ has
a weakly singular kernel, see~(\ref{eq:STCB}). For $c=1$, the
system~(\ref{eq:HKsys}) is identical to the ``KM2 system'' suggested
by Kleinman and Martin \cite[Eq.~(4.10)]{KleiMart88}, and further
discussed in \cite[Section~3.4]{HelsKarl18}. In
\cite[Theorem~4.3]{KleiMart88} it is proven that for $c=1$ the
system~(\ref{eq:HKsys}) has a unique solution if $\Gamma$ is smooth
and $k_1$ and $k_2$ both are real and positive.

To obtain uniqueness in~(\ref{eq:HKsys}) also for complex $k_1$ and
$k_2$ we let
\begin{equation}
\arg(c)=\left\{
\begin{array}{lll}
\arg(\varepsilon_2 k_2/\varepsilon_1)    &\mbox{if}&\Re{\rm e}\{k_1\}\ge 0\,,\\
\arg(\varepsilon_2 k_2/\varepsilon_1)-\pi&\mbox{if}&\Re{\rm e}\{k_1\}< 0\,.
\end{array}
\right.
\label{eq:cchoice}
\end{equation}
With such choices of $c$ one can show that \eqref{eq:HKsys} has a
unique solution if $\Gamma$ is smooth and
\begin{equation}
 0\le\arg(k_1),\arg(k_2)<\pi\,,
\quad 
\lvert\varepsilon_2/\varepsilon_1\rvert\ne\infty\,,
\quad 0\le\arg(\varepsilon_2 k_1/\varepsilon_1)\le\pi\,.
\label{eq:Klein}
\end{equation}
The conditions~(\ref{eq:Klein}) also guarantee that the transmission
problem of Section~\ref{sec:PDE} has a unique solution $U$ for
$N=2$~\cite[Section~4.1]{KleiMart88}.

To prove our claims about unique solvability for~(\ref{eq:HKsys}),
when $c$ obeys~(\ref{eq:cchoice}) we look at the system matrix in
\eqref{eq:HKsys}, whose real adjoint is
\begin{equation}
\begin{bmatrix}
I+\breve{\alpha}_1\left(\varepsilon_1^{-1} K^{\rm A}_1
+c\varepsilon_2^{-1}K^{\rm A}_2\right)&
\beta_1(T_1+T_2)\\
-\breve{\alpha}_1\left(S_1+cS_2\right)&
I-\beta_1\left(\varepsilon_1K_1+\varepsilon_2K_2\right)
\end{bmatrix}.
\label{eq:HKsysarrayadj}
\end{equation}
Applying a similarity transformation to \eqref{eq:HKsysarrayadj}
using the change-of-basis
\begin{equation}
\begin{bmatrix}
0 & I\\
\breve{\alpha}_1\varepsilon_1^{-1}\beta_1^{-1} I & 0
\end{bmatrix}
\end{equation}
gives the system block-matrix
\begin{equation}
\begin{bmatrix}
I-\beta_1\left(\varepsilon_1K_1+\varepsilon_2K_2\right)&
 -\beta_1\varepsilon_1\left(S_1+cS_2\right)\\
\breve{\alpha}_1\varepsilon_1^{-1}(T_1+T_2)&
I+\breve{\alpha}_1\left(\varepsilon_1^{-1} K^{\rm A}_1
+c\varepsilon_2^{-1}K^{\rm A}_2\right) 
\end{bmatrix}.
\label{eq:HKsysarrayadj2}
\end{equation}
Now \eqref{eq:HKsysarrayadj2} is identical to the system matrix of the
``KM1 system''~\cite[Eq.~(25)]{HelsKarl18}. Therefore, the analysis of
unique solvability of~(\ref{eq:HKsys}) and
of~\cite[Eq.~(25)]{HelsKarl18} are the same. In
\cite[Section~5.2]{HelsKarl18} it is shown, using
\cite[Theorem~4.1]{KleiMart88}, that~\cite[Eq.~(25)]{HelsKarl18} is
uniquely solvable on smooth $\Gamma$ whenever~(\ref{eq:Klein}) holds
and $c$ is chosen according to~(\ref{eq:cchoice}). The same then holds
for~(\ref{eq:HKsys}).

\medskip\noindent
{\bf Remark}: The system~\cite[Eq.~(25)]{HelsKarl18} is a special case
of~\cite[Eq.~(4.2)]{KleiMart88}. While the system~(\ref{eq:HKsys}) has
physical densities as unknowns, the
systems~\cite[Eq.~(25)]{HelsKarl18} and~\cite[Eq.~(4.2)]{KleiMart88}
have abstract densities as unknowns.

\subsection{Integral equations when $N>2$}

We now derive systems of integral equations with physical densities
for objects made up of more than one region, that is, for $N>2$.

For each subcurve $\Gamma_m$ we derive two integral equations. First
we add $(\varepsilon_m^{\rm R})^{-1}$ times \eqref{eq:ie1} for the
closed curve that bounds the region to the right of $\Gamma_m$ and
$(\varepsilon_m^{\rm L})^{-1}$ times \eqref{eq:ie1} for the closed
curve that bounds the region to the left of $\Gamma_m$. To this we add
$\varepsilon_n^{-1}$ times the null fields \eqref{eq:rep2} from all
other closed curves ${\cal C}_n$. The other equation is the sum of
\eqref{eq:ie2} for the two closed curves having $\Gamma_m$ in common.
To this we add the null fields \eqref{eq:rep4} from all other closed
curves ${\cal C}_n$. On each $\Gamma_m$ the integral equations then
read
\begin{equation}
\begin{bmatrix}
I+\alpha_m\sum\limits_{n=1}^N\varepsilon_n^{-1}K_n&
 -\alpha_m\sum\limits_{n=1}^NS_n\\
  \beta_m\sum\limits_{n=1}^N T_n&
I-\beta_m\sum\limits_{n=1}^N\varepsilon_nK_n^{\rm A}
\end{bmatrix}
\begin{bmatrix}
\mu(r)\\
\rho(r)
\end{bmatrix}
=
\begin{bmatrix}
f_{1m}(r)\\
f_{2m}(r)
\end{bmatrix},\quad r\in \Gamma_m,
\label{eq:inttotmat}
\end{equation}
where
\begin{gather}
\alpha_m=\frac{\varepsilon_m^{\rm R}\varepsilon_m^{\rm L}}
{\varepsilon_m^{\rm R}+\varepsilon_m^{\rm L}}\,,\quad 
\beta_m=\frac{1}{\varepsilon_m^{\rm R}+\varepsilon_m^{\rm L}}\,,
\label{eq:constants1}\\
f_{1m}(r)=2\alpha_m\varepsilon_1^{-1}U^{\rm in}(r)\,,\quad 
f_{2m}(r)=2\beta_m\nu\cdot\nabla U^{\rm in}(r)\,,
\label{eq:konstanter2}
\end{gather}
and where the global integral operators in the blocks
of~(\ref{eq:inttotmat}) have the same compactness properties as the
global operators in the corresponding blocks of~(\ref{eq:HKsys}).

If $U$ is a solution to the transmission problem of
Section~\ref{sec:PDE}, then $\mu$ and $\rho$ of~(\ref{eq:mu}),
(\ref{eq:rhoR}), and (\ref{eq:rhoL}) solve~(\ref{eq:ie1})
and~(\ref{eq:ie2}) and satisfy the null-field
representations~(\ref{eq:rep2}) and~(\ref{eq:rep4}). Therefore these
$\mu$ and $\rho$ also solve~(\ref{eq:inttotmat}). Unfortunately, we
are not able to prove unique solvability of~(\ref{eq:inttotmat}). For
$N=2$, however, the system~(\ref{eq:inttotmat}) reduces
to~(\ref{eq:HKsys}) with $c=1$. In view of Section~\ref{sec:unique}
one can therefore speculate that~(\ref{eq:inttotmat}) may be
particularly appropriate when all wavenumbers are real and positive.
The numerical examples of Section~\ref{sec:numex}, below, support this
view.

\section{Evaluation of electromagnetic fields}
\label{sec:eval}

The magnetic and electric fields $\myvec H$ and $\myvec E$ can be
obtained from $U$ via~(\ref{eq:H}) and~(\ref{eq:E}). The integral
representation of $U$ is \eqref{eq:rep1}.

\subsection{The magnetic field}

To \eqref{eq:rep1} we can add the null fields \eqref{eq:rep2} of
$\Omega_p$, $p\neq n$, and then
\begin{equation}
U(r)=U^{\rm in}(r)
-\frac{1}{2}\sum_{n=1}^N\left(K_n\mu(r)-\varepsilon_nS_n\rho(r)\right)\,,
\quad r\in\mathbb{R}^2\,.
\label{eq:rep1sum}
\end{equation}
Note that the incident field $U^{\rm in}$ is present in all regions,
but is extinct in all regions except $\Omega_1$, by fields generated
by the surface densities.

In contrast to the local representation~(\ref{eq:rep1}) of $U(r)$, the
representation~(\ref{eq:rep1sum}) is global and valid for all
$r\in\mathbb{R}^2$. A nice feature of~(\ref{eq:rep1sum}) is that the
sum over the regional layer potentials $K_n\mu(r)$ cancels the leading
singularities in the kernels of the individual $K_n\mu(r)$, see the
discussion after~(\ref{eq:Koper2}). This is an advantage when $U(r)$
is to be evaluated at $r$ very close to $\Gamma$.

We emphasize that the global representation of $U(r)$
in~\cite[Section~3]{GreeLee12} and~\cite[Section~3.2]{JHPAT17},
expressed in abstract densities, differs from our global
representation~(\ref{eq:rep1sum}) in several ways. For example, the
wavenumbers in the representations are determined according to
different criteria. Further, and more importantly, the global
representation in~\cite{GreeLee12,JHPAT17} does not lead to
kernel-singularity cancellation for $r$ close to $\Gamma$.

\subsection{The electric field}

The electric field is the vector field $\myvec E=E_x\hat{\myvec
  x}+E_y\hat{\myvec y}$, where
\begin{equation}
\begin{split}
E_x&= \frac{\rm i}{2k_0\varepsilon(r)}\hat{\myvec y}\cdot\nabla_3 U(r)\,,\\
E_y&=-\frac{\rm i}{2k_0\varepsilon(r)}\hat{\myvec x}\cdot\nabla_3 U(r)\,.
\end{split}
\label{eq:ExEy}
\end{equation} 
To find computable expressions for $E_x$ and $E_y$ we let $\nu=(0,1)$
and $\nu=(1,0)$ in \eqref{eq:rep3} and insert the resulting
expressions into \eqref{eq:ExEy}
\begin{equation}
\begin{split}
E_x(r)=&E_x^{\rm in}(r)\delta_{n1}
-\frac{\rm i}{2k_0\varepsilon_n}
\left(\partial_y K_n\mu(r)-\varepsilon_n\partial_yS_n\rho(r)\right), 
\quad r\in \Omega_n\,,\\
E_y(r)=&E_y^{\rm in}(r)\delta_{n1}
+\frac{\rm i}{2k_0\varepsilon_n}
\left(\partial_x K_n\mu(r)-\varepsilon_n\partial_xS_n\rho(r)\right), 
\quad r\in \Omega_n\,.
\end{split}
\label{eq:Erepxy}
\end{equation}
The near hypersingularities of $\partial_x K_n$ and $\partial_y K_n$
may destroy the numerical accuracy for $r$ close to ${\cal C}_n$. To
prevent this, the null fields of \eqref{eq:rep4} are added to
\eqref{eq:Erepxy} to obtain the global representation
\begin{equation}
\begin{split}
E_x(r)=&\dfrac{\varepsilon_1}{\varepsilon(r)}E_x^{\rm in}(r)
-\dfrac{\rm i}{2k_0\varepsilon(r)}\sum_{n=1}^N
\left(\partial_y K_n\mu(r)-\varepsilon_n\partial_yS_n\rho(r)\right),
\,\,\, r\in \mathbb{R}^2,\\
E_y(r)=&\dfrac{\varepsilon_1}{\varepsilon(r)}E_y^{\rm in}(r)
+\dfrac{\rm i}{2k_0\varepsilon(r)}\sum_{n=1}^N
\left(\partial_x K_n\mu(r)-\varepsilon_n\partial_xS_n\rho(r)\right),
\,\,\, r\in \mathbb{R}^2.
\end{split}
\label{eq:ErepxySUM}
\end{equation}

\section{An extended formulation for $\myvec E$}
\label{sec:evalextend}

When the ratio of wavenumbers (contrast) between regions is high, also
the representation~(\ref{eq:ErepxySUM}) of $\myvec E$ has some
problems to deliver high accuracy for $r$ close to $\Gamma$. The
alternative representation of $\myvec E$, that we now present, takes
care of this problem.

On each $\Gamma_m$ we introduce the electric surface charge density
$\varrho_{\rm E}$, the electric surface current density $\myvec J_{\rm
  s}$, and the magnetic surface current density $\myvec M_{\rm s}$ as
\begin{align}
\varrho_{\rm E}(r)&=\varepsilon_m^{\rm R}\myvec\nu\cdot
                    \myvec E^{\rm R}(r)\,,
\label{eq:varrhoE}\\
\myvec J_{\rm s}(r)&=\myvec\nu\times \myvec H(r)\,,\\  
\myvec M_{\rm s}(r)&=\myvec E(r)\times\myvec\nu\,,
\end{align}
where $\myvec E^{\rm R}$ is the limit of $\myvec E$ on the right-hand
side of $\Gamma_m$. This choice of surface densities is based on that
the tangential components of the magnetic and electric fields, and the
normal component of the electric flux density, $\varepsilon(r)\myvec
E(r)$, are continuous at all boundaries. The densities $\myvec J_{\rm
  s}$ and $\myvec M_{\rm s}$ are expressed in the densities $\mu$ and
$\rho$ as
\begin{align}
\myvec J_{\rm s}(r)&=-\myvec\tau\mu(r)\,,
\label{eq:JsMs1}\\
\myvec M_{\rm s}(r)&={\rm i}k_0^{-1}\hat{\myvec z}\rho(r)\,,
\label{eq:JsMs2}
\end{align}
and are by that known once \eqref{eq:inttotmat} is solved. 

The alternative integral representation of $\myvec E$ is
\begin{multline}
\myvec E(r)=\myvec E^{\rm inc}(r)\delta_{n1}
+\dfrac{1}{2}\varepsilon_n^{-1}\nabla_3 S_n\varrho_{\rm E}(r)\\
-\dfrac{1}{2}{\rm i}k_0^{-1}\hat{\myvec z}\times\nabla_3 S_n\rho(r)
+\dfrac{1}{2}{\rm i}k_0\myvec B_n\mu(r)\,, \quad r\in\Omega_n\,.
\label{eq:ErepMOD}
\end{multline}
This is the two-dimensional equivalent of the integral representation
of $\myvec E$ derived, using a vector analogue of Green's theorem, in
\cite{Kris16,Muller69,Strom75} and given by \cite[Eq.~(3.12), upper
line]{Kris16}, \cite[page~132,~Eq.~(16)]{Muller69}, and
\cite[Eq.~(2.10), upper line]{Strom75}. It is also possible to
derive~(\ref{eq:ErepMOD}) from~(\ref{eq:rep1}) by multiplication
of~(\ref{eq:rep1}) with $\hat{\myvec z}$, application of the curl
operator, and integration by parts. The null-field representation
accompanying~(\ref{eq:ErepMOD}) is, see \cite[Eq.~(2.10), lower
line]{Strom75},
\begin{multline}
\myvec 0=\myvec E^{\rm inc}(r)\delta_{n1}
+\dfrac{1}{2}\varepsilon_n^{-1}\nabla_3 S_n\varrho_{\rm E}(r)\\
-\dfrac{1}{2}{\rm i}k_0^{-1}\hat{\myvec z}\times\nabla_3 S_n\rho(r)
+\dfrac{1}{2}{\rm i}k_0\myvec B_n\mu(r)\,, \quad r\notin\Omega_n\,.
\label{eq:nullfE}
\end{multline}

From~(\ref{eq:varrhoE}), (\ref{eq:ErepMOD}), and (\ref{eq:nullfE}) an
integral equation for $\varrho_{\rm E}$ can be found as
\begin{multline}
\varrho_{\rm E}(r)
-\alpha_m\sum_{n=1}^N\varepsilon_n^{-1}K_n^{\rm A}\varrho_{\rm E}(r)=
 2\myvec\nu\cdot\myvec E^{\rm in}(r)\\
+\alpha_m{\rm i}k_0^{-1}\sum_{n=1}^NC_n\rho(r)
+\alpha_m{\rm i}k_0\myvec\nu\cdot\sum_{n=1}^N\myvec B_n\mu(r)\,,
\quad r\in\Gamma_m\,.
\label{eq:varhosum}
\end{multline}
Here $\alpha_m$ is given in \eqref{eq:constants1}, the global integral
operator on the left hand side is compact away from singular boundary
points, and the global integral operators on the right hand side are
everywhere compact.

We remark that solvers for Maxwell transmission problems based on
solving Müller-type equations, such as three-dimensional counterparts
of~(\ref{eq:inttotmat}), augmented with integral equations for
current- and charge densities, such as~(\ref{eq:varhosum}), are
referred to as {\it charge-current formulations}~\cite{VicGreFer18}.

We also remark that the singular nature of the kernels of the layer
potentials in the global representation~(\ref{eq:ErepxySUM}) of
$\myvec E$ and the local representation~(\ref{eq:ErepMOD}) of $\myvec
E$ are similar. Both representations have kernels that exhibit
Cauchy-type singularities as $\mathbb{R}^2\setminus\Gamma\ni r\to
r'\in\Gamma$, see Section~\ref{sec:singnat}.

\section{Separated curves and distant regions}
\label{sec:sepdist}

We say that a closed curve ${\cal C}_n$ is separated from a subcurve
$\Gamma_m$ if ${\cal C}_n$ and $\Gamma_m$ have no common points. We
say that a closed curve ${\cal C}_n$ is distant to a point $r$ if
$r\notin \Omega_n$ and if ${\cal C}_n$ is far enough from $r$ to make
singularities in the kernels of the regional layer potentials
$G_n\sigma(r)$ of~(\ref{eq:Gopern}) harmless from a numerical point of
view.

Surface densities on a curve ${\cal C}_n$ that is separated from a
subcurve $\Gamma_m$ do not contribute with null fields on $\Gamma_m$
that cancel individual kernel singularities in the integral equations
\eqref{eq:inttotmat} and \eqref{eq:varhosum}. This means that the
corresponding terms can be excluded from the sums in
\eqref{eq:inttotmat} and \eqref{eq:varhosum}. The same applies to
curves that are distant to $r$ in the global representations
\eqref{eq:rep1sum} and \eqref{eq:ErepxySUM}.

As an example consider the object in Figure \ref{fig:amoeba2}. Here
${\cal C}_3$ is separated from $\Gamma_4$ and $\Gamma_6$ and ${\cal
  C}_4$ is separated from $\Gamma_2$ and $\Gamma_5$. The curve ${\cal
  C}_3$ is distant to $r\in\Omega_4$ and ${\cal C}_4$ is distant to
$r\in\Omega_3$.

\section{Discretization}
\label{sec:disc}

We discretize and solve our systems of integral equations using
Nyström discretization with composite $16$-point Gauss--Legendre
quadrature as underlying quadrature. Starting from a coarse uniform
mesh on $\Gamma$, extensive temporary dyadic mesh refinement is
carried out in directions toward corners and triple junctions.
Explicit kernel-split-based product integration is used for
discretization of singular parts of operators. Recursively compressed
inverse preconditioning (RCIP) is used for lossless compression in
tandem with the mesh refinement so that the resulting linear system
has unknowns only on a grid on the coarse mesh. The final linear
system is then solved iteratively using GMRES. For field evaluations
near $\Gamma$ in post-processors we, again, resort to explicit
kernel-split product integration in order to accurately resolve near
singularities in layer-potential kernels.

The overall discretization scheme, summarized in the paragraph above,
is basically the same as the scheme used for Helmholtz transmission
problems with two non-smooth dielectric regions $\Omega_n$
in~\cite{HelsKarl18}, and we refer to that paper for details. The RCIP
technique, which can be viewed as a locally applicable fast direct
solver, has previously been used for integral equation reformulations
of transmission problems of other piecewise-constant-coefficient
elliptic PDEs in domains that involve region interfaces that meet at
triple junctions. See, for example,~\cite{HelsOjal09}. See also the
compendium~\cite{Hels18} for a thorough review of the RCIP technique.

\section{Comparison with other formulations for $N=2$}
\label{sec:compare}

Before venturing into numerical examples, we relate our new
system~(\ref{eq:HKsys}) and its real
adjoint~\cite[Eq.~(25)]{HelsKarl18} to a few popular integral equation
formulations for the Maxwell transmission problem in three
dimensions~\cite{LaiJian18,Muller69,VicGreFer18}, adapted to the
problem of Section~\ref{sec:PDE} with $N=2$. Recall
that~(\ref{eq:HKsys}) and~\cite[Eq.~(25)]{HelsKarl18} have unique
solutions when~(\ref{eq:Klein}) holds thanks to the
choice~(\ref{eq:cchoice}) of the parameter $c$. This allows for purely
negative ratios $\varepsilon_2/\varepsilon_1$ when $k_1$ is real and
positive. Do other formulations have unique solutions in this regime,
too? and, if not, can they be modified so that they do?

When the Müller system~\cite[p.~319]{Muller69} is adapted to the
problem of Section~\ref{sec:PDE}, it reduces to~(\ref{eq:HKsys}) with
$c=\varepsilon_2/\varepsilon_1$. This value is not compatible
with~(\ref{eq:cchoice}) and a unique solution can not be guaranteed
for negative $\varepsilon_2/\varepsilon_1$. Note the sign error
in~\cite[Eq.~(40), p.~301]{Muller69} which carries over
to~\cite[p.~319]{Muller69}, as observed in~\cite[p.~83]{Mautz77}.

The system of Lai and Jiang~\cite[Eqs.~(40)-(42)]{LaiJian18} is the
real adjoint of the Müller system. When adapted to the problem of
Section~\ref{sec:PDE}, and with use of partial integration in Maue's
identity in two dimensions~\cite[Eq.~(2.4)]{Kres95}, the system matrix
reduces to~(\ref{eq:HKsysarrayadj2}) with
$c=\varepsilon_2/\varepsilon_1$ and, again, invertibility can not be
guaranteed for negative $\varepsilon_2/\varepsilon_1$.

The ``$\myvec H$-system'' of Vico, Greengard, and Ferrando, coming
from the representation $\myvec H$ of~\cite[Eq.~(38)]{VicGreFer18},
reduces to the same system as that to
which~\cite[Eqs.~(40)-(42)]{LaiJian18} reduces to, when adapted to the
problem of Section~\ref{sec:PDE}. The conclusion about unique
solvability is the same.

In an attempt to modify the ``$\myvec H$-system''
of~\cite{VicGreFer18} so that it becomes uniquely solvable also for
negative $\varepsilon_2/\varepsilon_1$, we introduce a parameter $c$
in the representation of $\myvec H$~\cite[Eq.~(38)]{VicGreFer18}
\begin{equation}
\begin{split}
\myvec H&=\varepsilon_1\nabla\times S_{k_1}[\myvec a]
         -\varepsilon_1S_{k_1}[\myvec n\sigma]
         +\varepsilon_1S_{k_1}[\myvec b]
         +\nabla S_{k_1}[\rho],
          \quad\myvec r\in\mathbb{R}^3\setminus\overline{D}\,,\\
\myvec H&=\varepsilon_2\nabla\times S_{k_2}[\myvec a]
         -\varepsilon_2S_{k_2}[\myvec n\sigma]
  +c\varepsilon_1(S_{k_2}[\myvec b]+\varepsilon_2^{-1}\nabla S_{k_2}[\rho]),
          \quad \myvec r\in D\,.
\end{split}
\label{eq:vico12}
\end{equation}
Here $D$, $S_k$ and $\nabla$ denote the object, the acoustic single
layer operator, and the nabla-operator in three dimensions. The choice
$c=\varepsilon_2/\varepsilon_1$ in~(\ref{eq:vico12}) leads to the
``$\myvec H$-system'' of~\cite{VicGreFer18}. It is probably better to
choose $c$ as in~(\ref{eq:cchoice}) since this leads to a system
which, when adapted to the problem of Section~\ref{sec:PDE}, reduces
to the uniquely solvable system~\cite[Eq.~(25)]{HelsKarl18}.

Similarly, we introduce $c$ also in the representation of $\myvec
E$~\cite[Eq.~(36)]{VicGreFer18}
\begin{equation}
\begin{split}
\myvec E&=\nabla\times S_{k_1}[\myvec a]
         -S_{k_1}[\myvec n\sigma]
         +\varepsilon_1S_{k_1}[\myvec b]
     +\nabla S_{k_1}[\rho]\,, 
      \quad\myvec r\in\mathbb{R}^3\setminus\overline{D}\,,\\
\myvec E&=\nabla\times S_{k_2}[\myvec a]
         -S_{k_2}[\myvec n\sigma]
         +c\varepsilon_1(S_{k_2}[\myvec b]
          +\varepsilon_2^{-1}\nabla S_{k_2}[\rho]), 
           \quad\myvec r\in D\,.
\end{split}
\label{eq:vico23}
\end{equation}
The choice $c=\varepsilon_2/\varepsilon_1$ leads to the ``$\myvec
E$-system''~\cite[Eq.~(37)]{VicGreFer18} which, when adapted to the
problem of Section~\ref{sec:PDE} and according to numerical
experiments, does not guarantee unique solutions for negative
$\varepsilon_2/\varepsilon_1$. If, on the other hand, $c$
in~(\ref{eq:vico23}) is chosen as in~(\ref{eq:cchoice}), then the
corresponding system appears to be uniquely solvable. See, further,
Section~\ref{sec:uniquex}.

\section{Numerical examples}
\label{sec:numex}

In three numerical examples, chosen as to resemble examples previously
treated in the literature, we now put our systems of integral
equations for $\mu$ and $\rho$ and our field representations of $U$
and $\myvec E$ to the test. When assessing the accuracy of computed
quantities we adopt a procedure where to each numerical solution we
also compute an overresolved reference solution, using roughly 50\%
more points in the discretization of the integral equations. The
absolute difference between these two solutions is denoted the {\it
  estimated absolute error}.

Our codes are implemented in {\sc Matlab}, release 2016b, and executed
on a workstation equipped with an Intel Core i7-3930K CPU. The
implementations are chiefly standard, rely on built-in functions, and
include a few {\tt parfor}-loops (which execute in parallel). Large
linear systems are solved using GMRES, incorporating a low-threshold
stagnation avoiding technique applicable to systems coming from
discretizations of Fredholm integral equations of the second
kind~\cite[Section~8]{HelsOjal08}. The GMRES stopping criterion is set
to machine epsilon in the estimated relative residual.

\subsection{A three-region example}

\begin{figure}[t]
\centering
\includegraphics[height=38mm, trim=0mm -20mm 0mm 20mm]{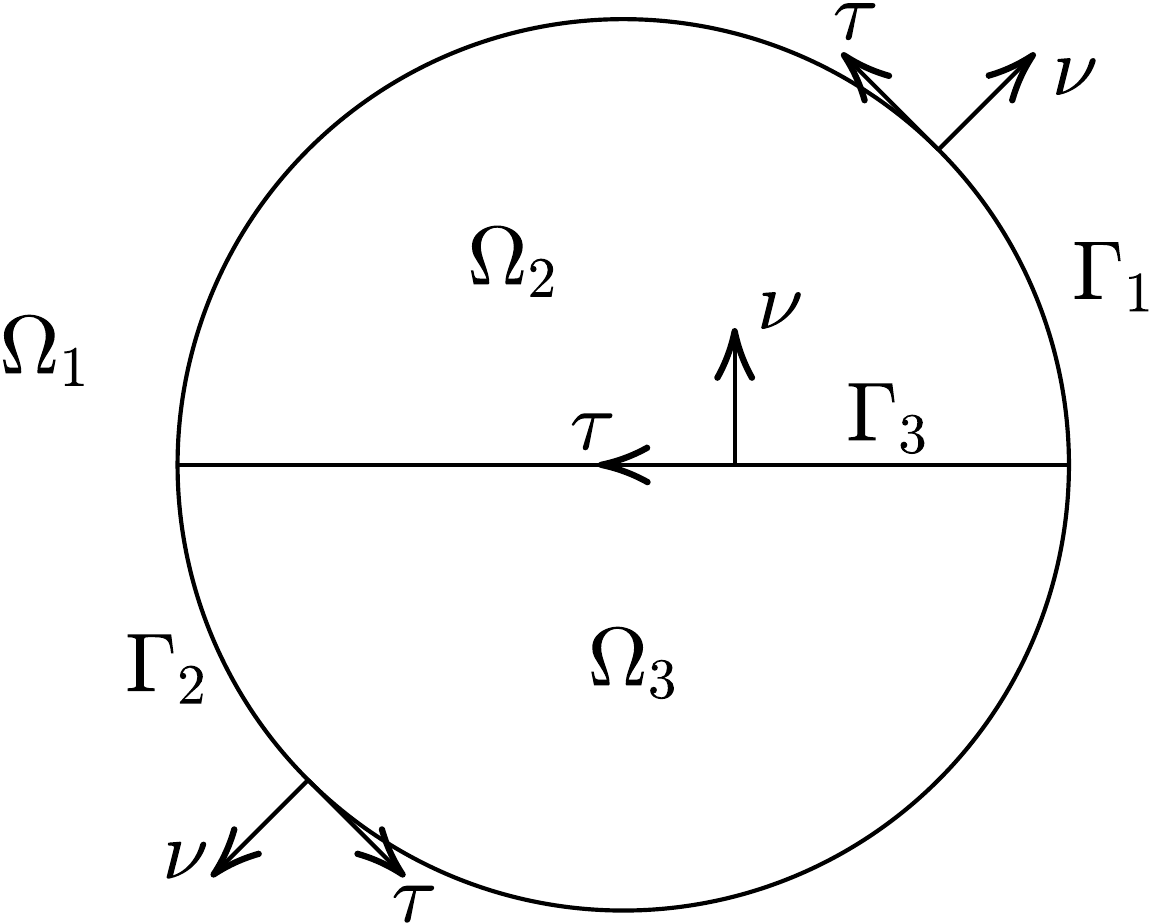}
\hspace*{5mm}
\includegraphics[height=50mm]{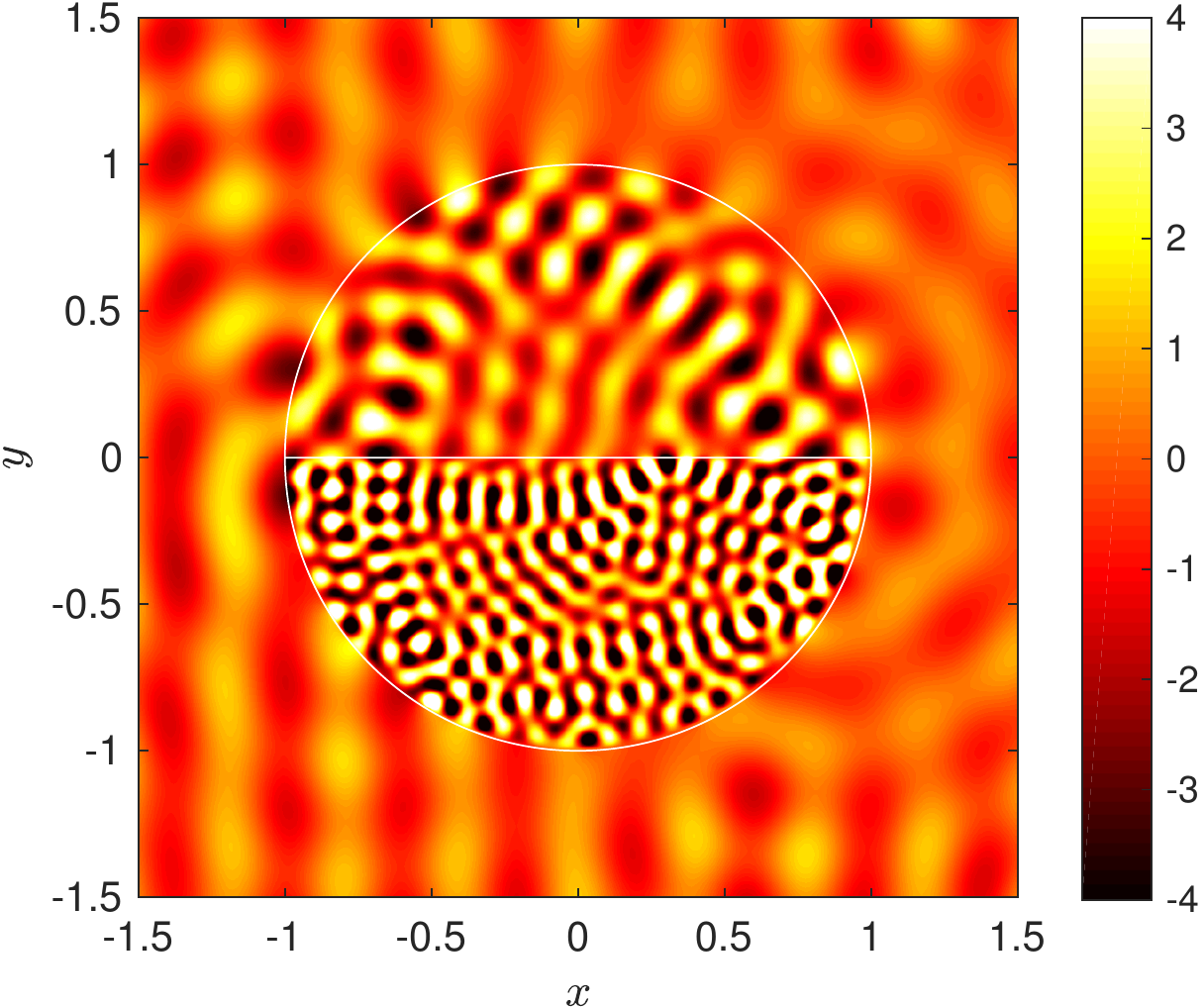}
\caption{\sf A three-region example. The vacuum wavenumber is 
  $k_0=16$ and the relative permittivities are $\varepsilon_1=1$,
  $\varepsilon_2=4$, and $\varepsilon_3=16$. The incident field
  travels in the direction $d=(1,0)$. Left: the configuration. Right:
  $H_z(r,0)$ with colormap ``hot'' and a colorbar range restricted to
  $[-4,4]$, as in~\cite[Figure~4(a)]{JHPAT17}.}
\label{fig:amoeba1}
\end{figure}

We start with a three-region example from Jerez-Hanckes,
Pérez-Aran\-ci\-bia, and Turc~\cite[Figure~4(a)]{JHPAT17}. The bounded
object is a unit disk divided into two equisized regions, see
Figure~\ref{fig:amoeba1}. The vacuum wavenumber is $k_0=16$ and the
relative permittivities are $\varepsilon_1=1$, $\varepsilon_2=4$, and
$\varepsilon_3=16$. The incident field travels in the direction
$d=(1,0)$.

\subsubsection{Numerical results for $\myvec H$ and $\myvec E$}

\begin{figure}[t]
\centering
\includegraphics[height=50mm]{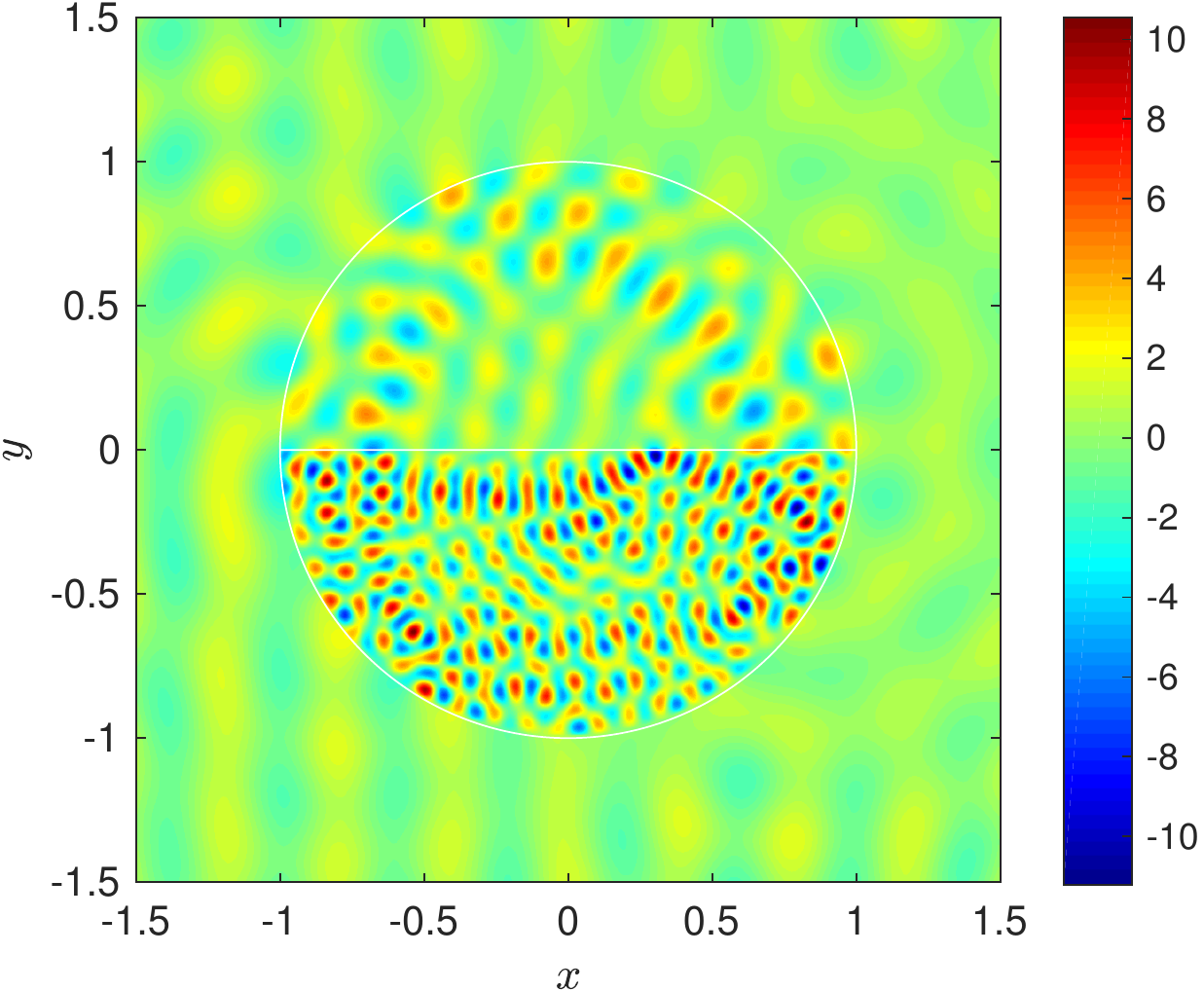}
\includegraphics[height=50mm]{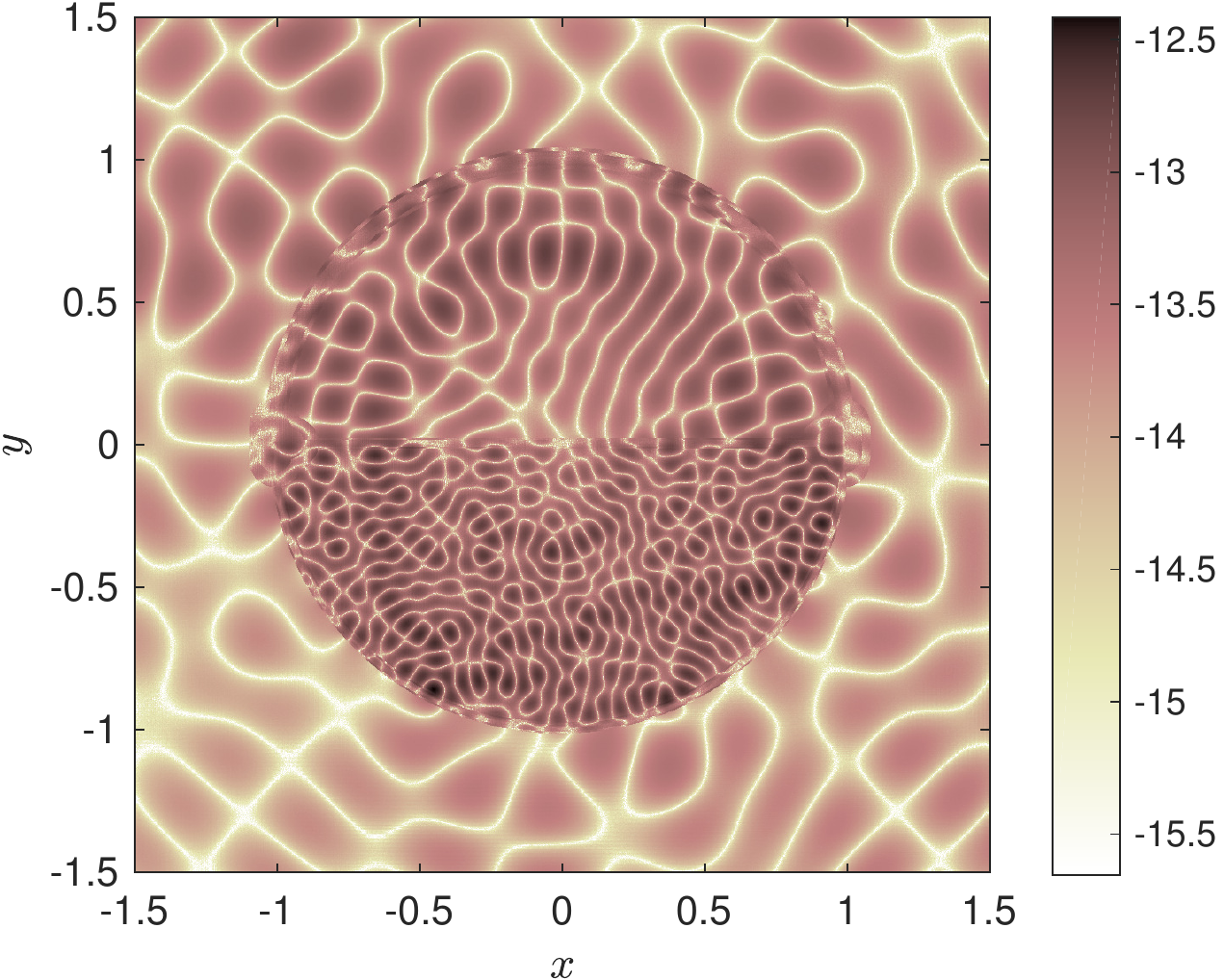}
\includegraphics[height=50mm]{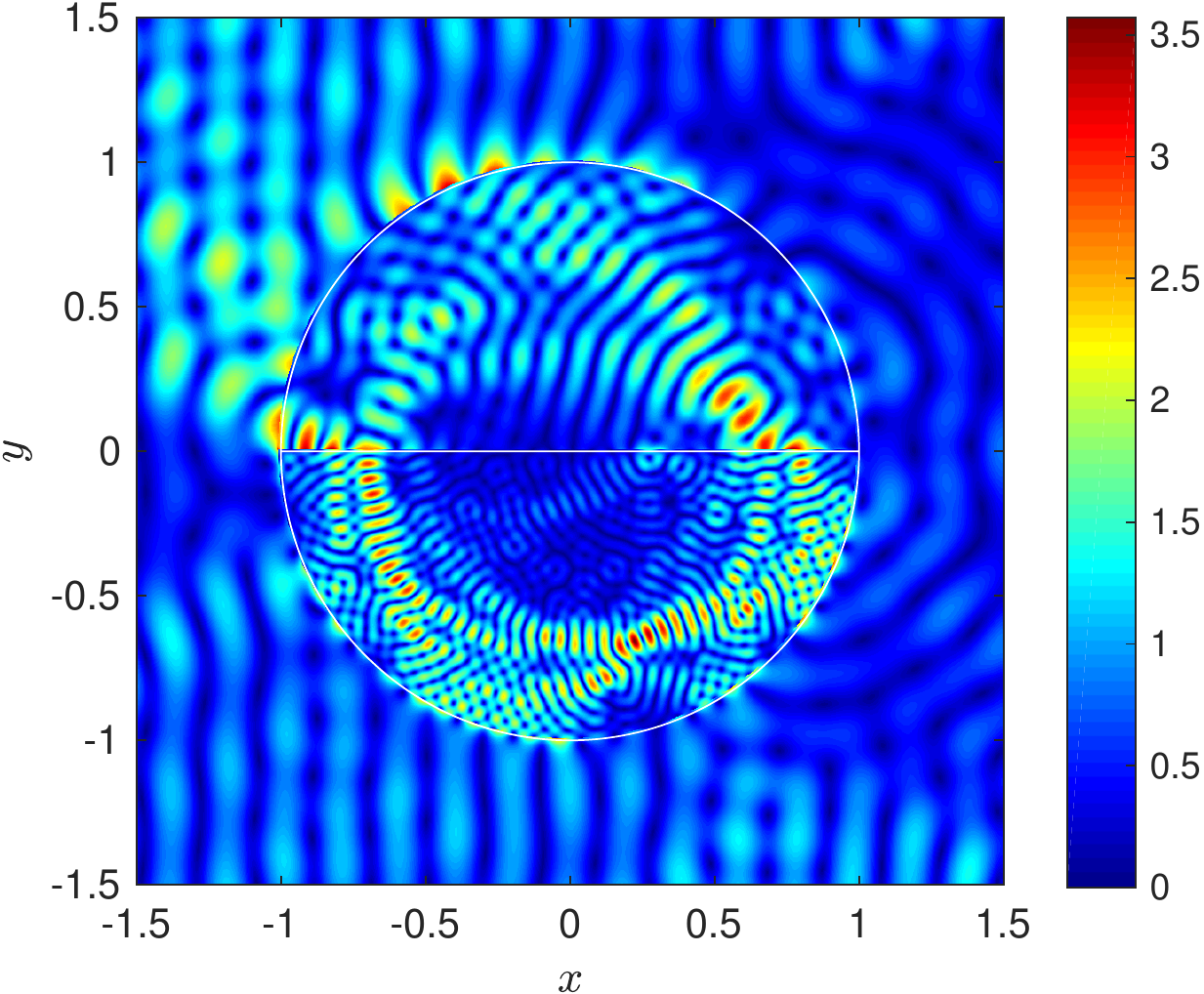}
\includegraphics[height=50mm]{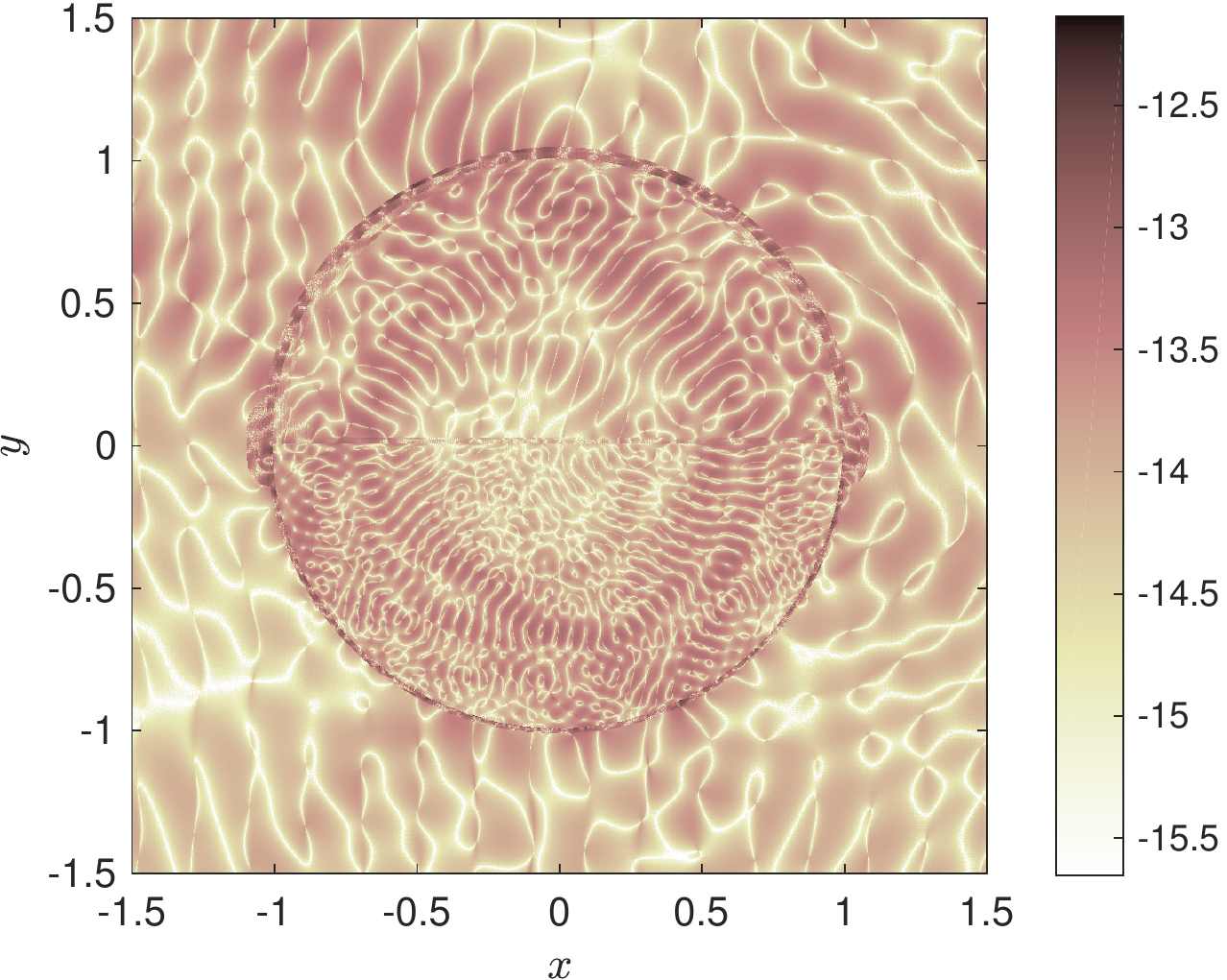}
\caption{\sf Numerical results for the three-region example in 
  Figure~\ref{fig:amoeba1}. Top left: the field $H_z(r,0)$. Top right:
  $\log_{10}$ of estimated absolute field error in $H_z(r,0)$. Bottom
  left: the field $\lvert\myvec E(r,0)\rvert$. Bottom right:
  $\log_{10}$ of estimated absolute field error in $\lvert\myvec
  E(r,0)\rvert$.}
\label{fig:bicircUE}
\end{figure}

The system~(\ref{eq:inttotmat}) is solved using $1,\!696$
discretization points on the coarse mesh on $\Gamma$. Results for
subsequent evaluations of the $z$-component of $\myvec H(r,0)$ and of
$\lvert \myvec E(r,0)\rvert$ are shown in Figure~\ref{fig:bicircUE}.
The local representations~(\ref{eq:rep1}) and~(\ref{eq:Erepxy}) are
used for field points $r$ away from $\Gamma$ and the global
representations~(\ref{eq:rep1sum}) and~(\ref{eq:ErepxySUM}) are used
for field points $r$ close to $\Gamma$. We quote the following
approximate timings: setting up the discretized
system~(\ref{eq:inttotmat}) took $7$ seconds, constructing various
quantities needed in the RCIP scheme took $50$ seconds, solving the
main linear system required $208$ GMRES iterations and took $3.5$
seconds. Computing $\myvec H$ and $\myvec E$ at $10^6$ field points
placed on a Cartesian grid in the box ${\cal B}=\left\{-1.5\le x\le
  1.5, -1.5\le y\le 1.5\right\}$ took, on average, $0.0011$ and
$0.0022$ seconds per point, respectively.

\subsubsection{Comparison with previous results for $\myvec H$}

The top left image of Figure~\ref{fig:bicircUE} shows the field
$H_z(r,0)$ using {\sc Matlab}'s colormap ``jet'' and a colorbar range
chosen as to include all values of $H_z(r,0)$ occurring in the box
${\cal B}$. For comparison with results
in~\cite[Figure~4(a)]{JHPAT17}, the right image of
Figure~\ref{fig:amoeba1} shows results with colormap ``hot'' and a
colorbar range restricted to $[-4,4]$. A close comparison between the
right image of Figure~\ref{fig:amoeba1}
and~\cite[Figure~4(a)]{JHPAT17} reveals that the figures look rather
similar, except for at field points $r$ very close to $\Gamma$, where
we think that our results are substantially more accurate than those
of~\cite[Figure~4(a)]{JHPAT17}.

\subsubsection{Results for $\myvec E$ via the extended representation}

For comparison we also compute $\myvec E(r,0)$ via the extended
representation~(\ref{eq:ErepMOD}) rather than via~(\ref{eq:Erepxy})
and~(\ref{eq:ErepxySUM}). This involves augmenting the system of
integral equations~(\ref{eq:inttotmat}) with the extra
equation~(\ref{eq:varhosum}). The estimated error in $\myvec E(r,0)$
(not shown) is slightly improved by switching to the extended
representation and resembles the error for $H_z(r,0)$ in the top right
image of Figure~\ref{fig:bicircUE}. The timings were affected as
follows: setting up the discretized augmented
system~(\ref{eq:inttotmat}) with~(\ref{eq:varhosum}) took $15$
seconds, constructing various quantities needed in the RCIP scheme
took $75$ seconds, solving the main linear system required $233$ GMRES
iterations and took $7$ seconds. Computing $\myvec E(r,0)$ at $10^6$
field points $r$ in the box ${\cal B}$ took, on average, $0.0013$
seconds per point. That is, the system setup and solution take longer
due to the extra unknown $\varrho_{\rm E}$, while the field
evaluations are faster since we use the local
representation~(\ref{eq:ErepMOD}) for all $r\in{\cal B}$ and avoid the
expensive global representation~(\ref{eq:ErepxySUM}) for $r$ close to
$\Gamma$.

\subsection{A four-region example}

We now turn our attention to the four-region configuration of
Figure~\ref{fig:amoeba2}. The object can be described as a unit disk
centered at the origin and with two smaller disks of half the radius
and origins at $x=\pm 1$ superimposed. The vacuum wavenumber is
$k_0=10$ and the relative permittivities are $\varepsilon_1=1$,
$\varepsilon_2=100$, and $\varepsilon_3=\varepsilon_4=625$. The
incident field travels in the direction $d=(1,0)$. The example is
inspired by the three-region high-contrast example
of~\cite[Figure~5]{GreeLee12}

\subsubsection{Numerical results for $\myvec H$}

\begin{figure}[t]
\centering
\includegraphics[height=37mm]{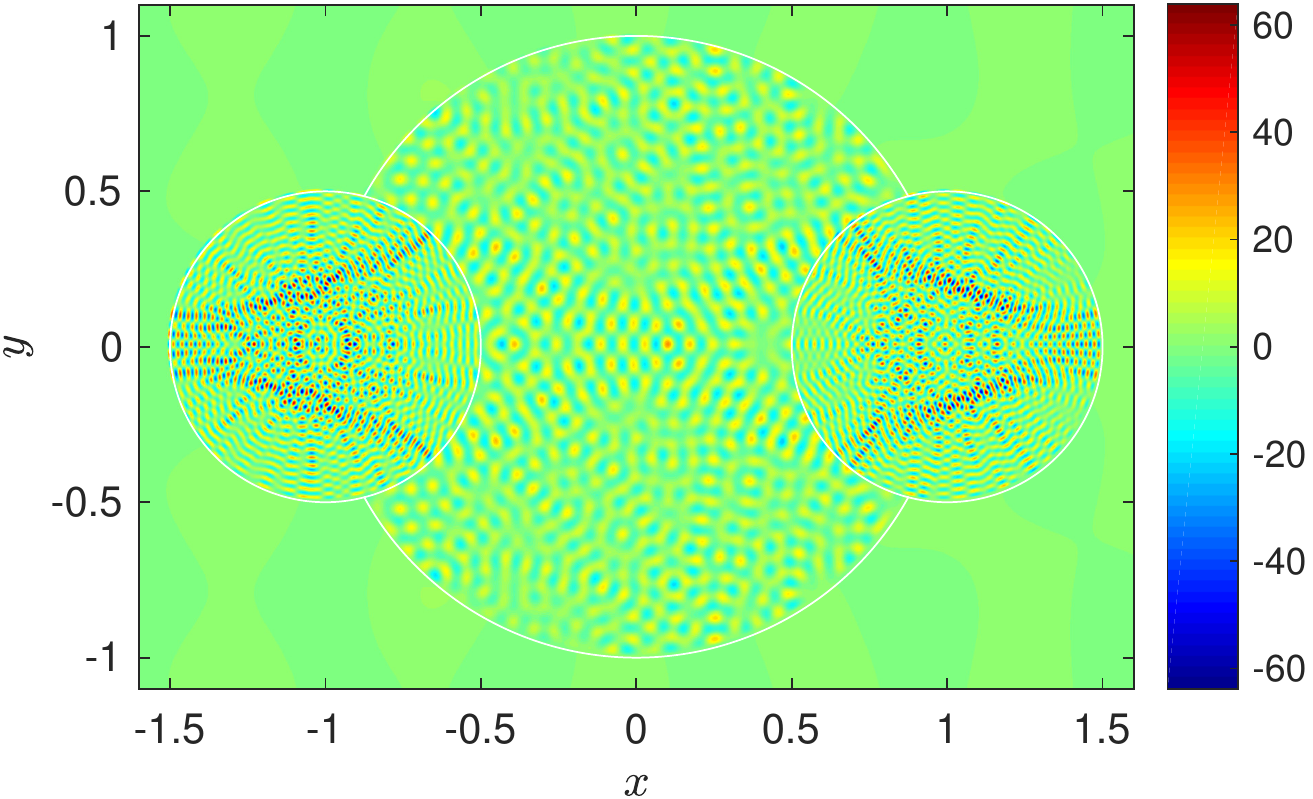}
\includegraphics[height=37mm]{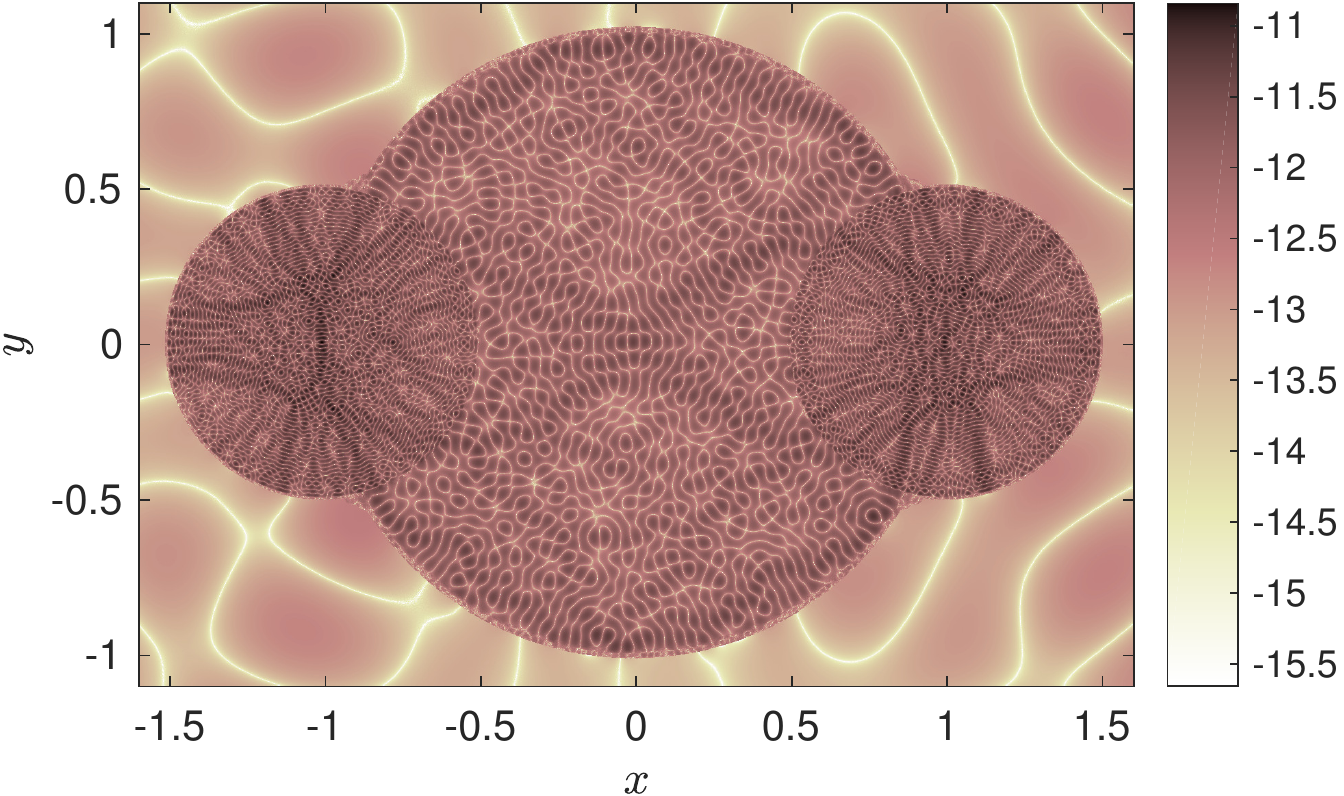}
\includegraphics[height=37mm]{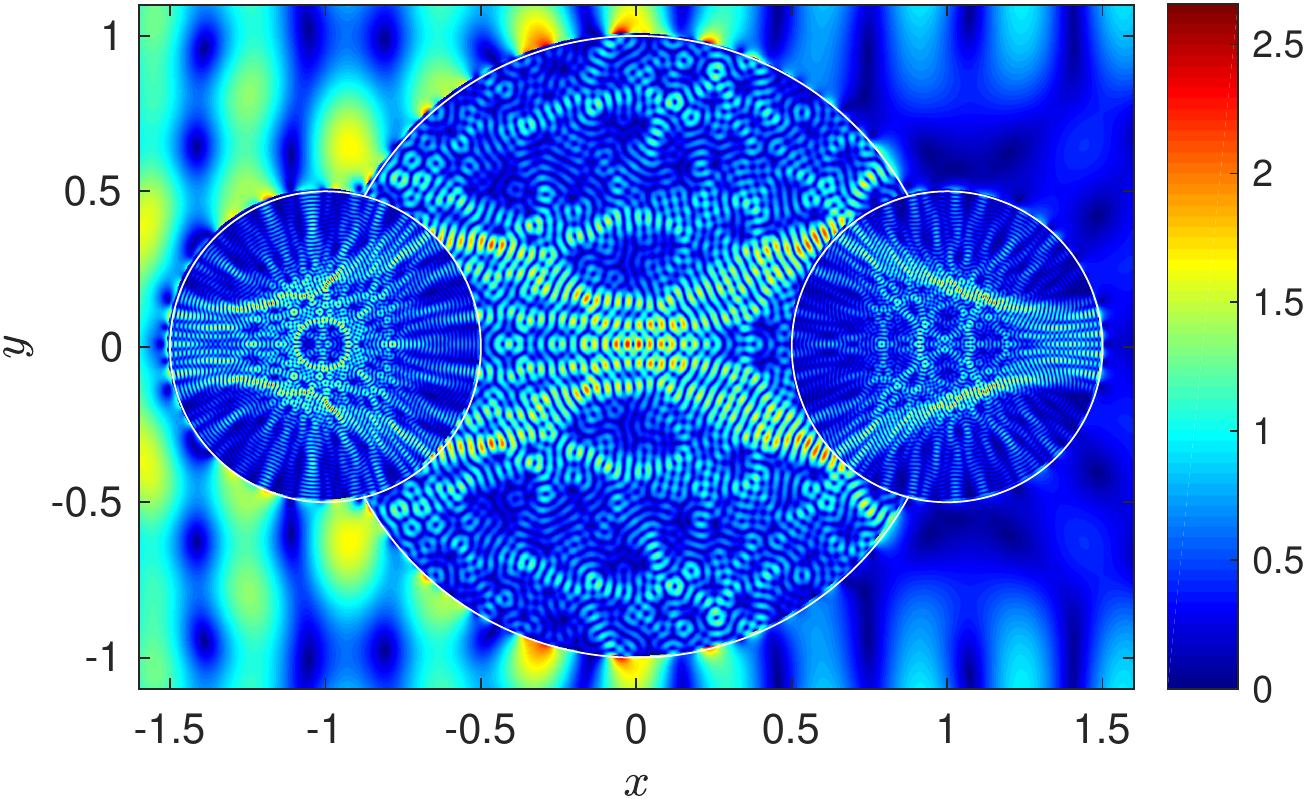}
\includegraphics[height=37mm]{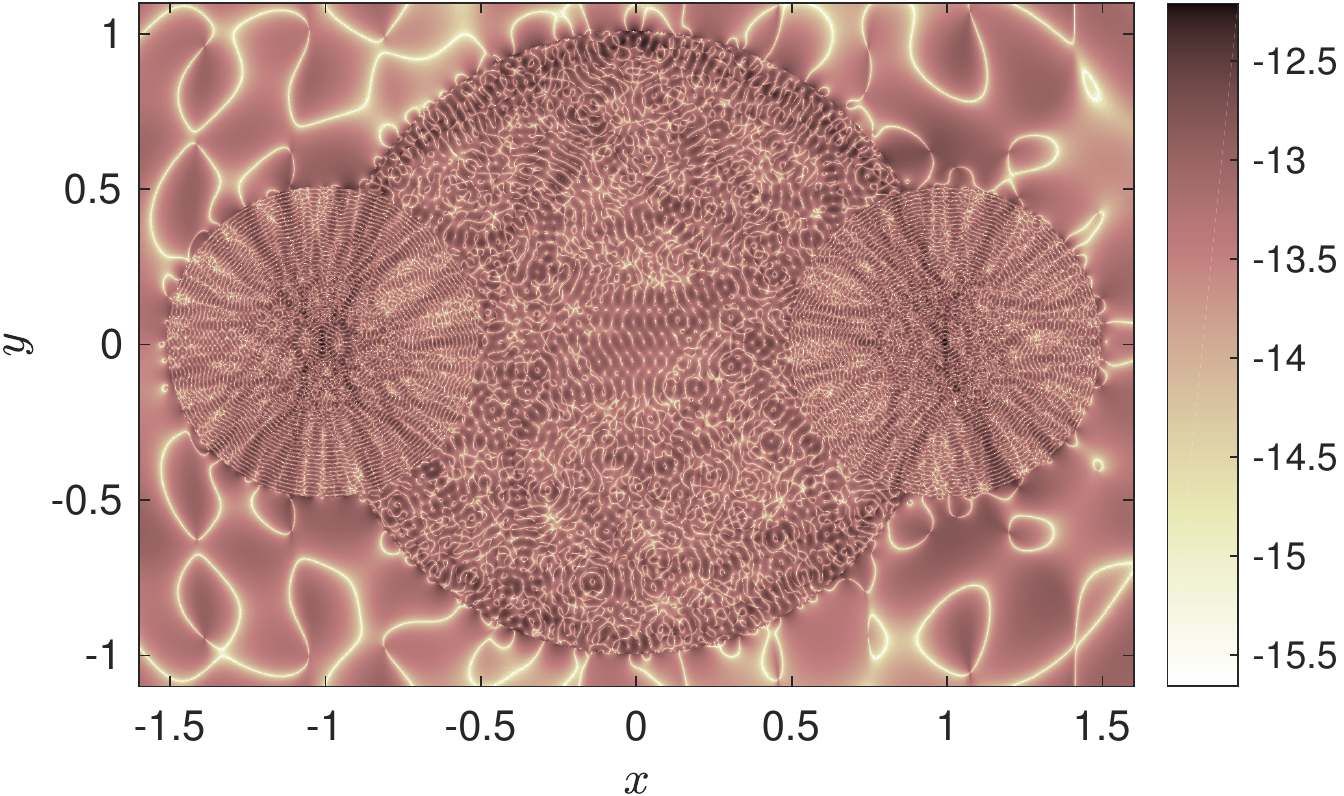}
\caption{\sf Results for the four-region example. The vacuum
  wavenumber is $k_0=10$ and the relative permittivities are
  $\varepsilon_1=1$, $\varepsilon_2=100$, and
  $\varepsilon_3=\varepsilon_4=625$. The incident field travels in the
  direction $d=(1,0)$. Top left: the field $H_z(r,0)$. Top right:
  $\log_{10}$ of estimated absolute field error in $H_z(r,0)$. Bottom
  left: the field $\lvert\myvec E(r,0)\rvert$. Bottom right:
  $\log_{10}$ of estimated absolute field error in $\lvert\myvec
  E(r,0)\rvert$.}
\label{fig:tricircUE}
\end{figure}

The magnetic field is computed by first discretizing and solving the
system~(\ref{eq:inttotmat}) using $5,\!600$ discretization points on
the coarse mesh on $\Gamma$ and then using~(\ref{eq:rep1})
or~(\ref{eq:rep1sum}) for field evaluations, depending on whether
field points $r$ are far away from $\Gamma$ or not.
In~(\ref{eq:inttotmat}), we exclude parts of operators corresponding
to (zero) contributions from surface densities on closed curves to
subcurves that are separated in the sense of
Section~\ref{sec:sepdist}. Similarly, in~(\ref{eq:rep1sum}), we
exclude (zero) field contributions from layer potentials on closed
curves that are distant to field points $r$. Numerical results are
shown in the top row of Figure~\ref{fig:tricircUE}. Timings are as
follows: setting up the discretized system~(\ref{eq:inttotmat}) took
$70$ seconds, constructing various quantities needed in the RCIP
scheme took $50$ seconds, solving the main linear system required
$596$ GMRES iterations and took $90$ seconds. Computing $\myvec H$ at
$1.455\times 10^6$ field points placed on a Cartesian grid in the box
${\cal B}=\left\{-1.6\le x\le 1.6, -1.1\le y\le 1.1\right\}$ took, on
average, $0.0028$ seconds per point.

\subsubsection{Numerical results for $\myvec E$}
\label{sec:fourE}

When the dielectric contrast between the regions is high, the extended
representation~(\ref{eq:ErepMOD}) of $\myvec E$ offers better accuracy
at field points $r$ very close to $\Gamma$ than do~(\ref{eq:Erepxy})
and~(\ref{eq:ErepxySUM}). Numerical results obtained
with~(\ref{eq:ErepMOD}) and with $7,\!008$ discretization points on
the coarse mesh on $\Gamma$ are shown in the bottom row of
Figure~\ref{fig:tricircUE}. Timings are as follows: setting up the
discretized system~(\ref{eq:inttotmat}) with~(\ref{eq:varhosum}) took
$125$ seconds, constructing various quantities needed in the RCIP
scheme took $75$ seconds, solving the main linear system required
$634$ GMRES iterations and took $300$ seconds. Computing $\myvec E$ at
the $1.455\times 10^6$ field points in ${\cal B}$ took, on average,
$0.0043$ seconds per point.

\subsubsection{A $20$ times triple-junction zoom for $\myvec E$}
\label{sec:triple}

We repeat the experiment of Section~\ref{sec:fourE}, zooming in on the
subcurve triple junction at
$\gamma_1=(-7/8,\sqrt{15}/8)\approx(-0.875,0.484)$ with a $20$ times
magnification. This means that we evaluate $\myvec E$ at $1.455\times
10^6$ field points in the box ${\cal B}\approx\left\{-0.955\le x\le
  -0.795, 0.429\le y\le 0.539\right\}$.

\begin{figure}[t]
\centering
\includegraphics[height=37mm]{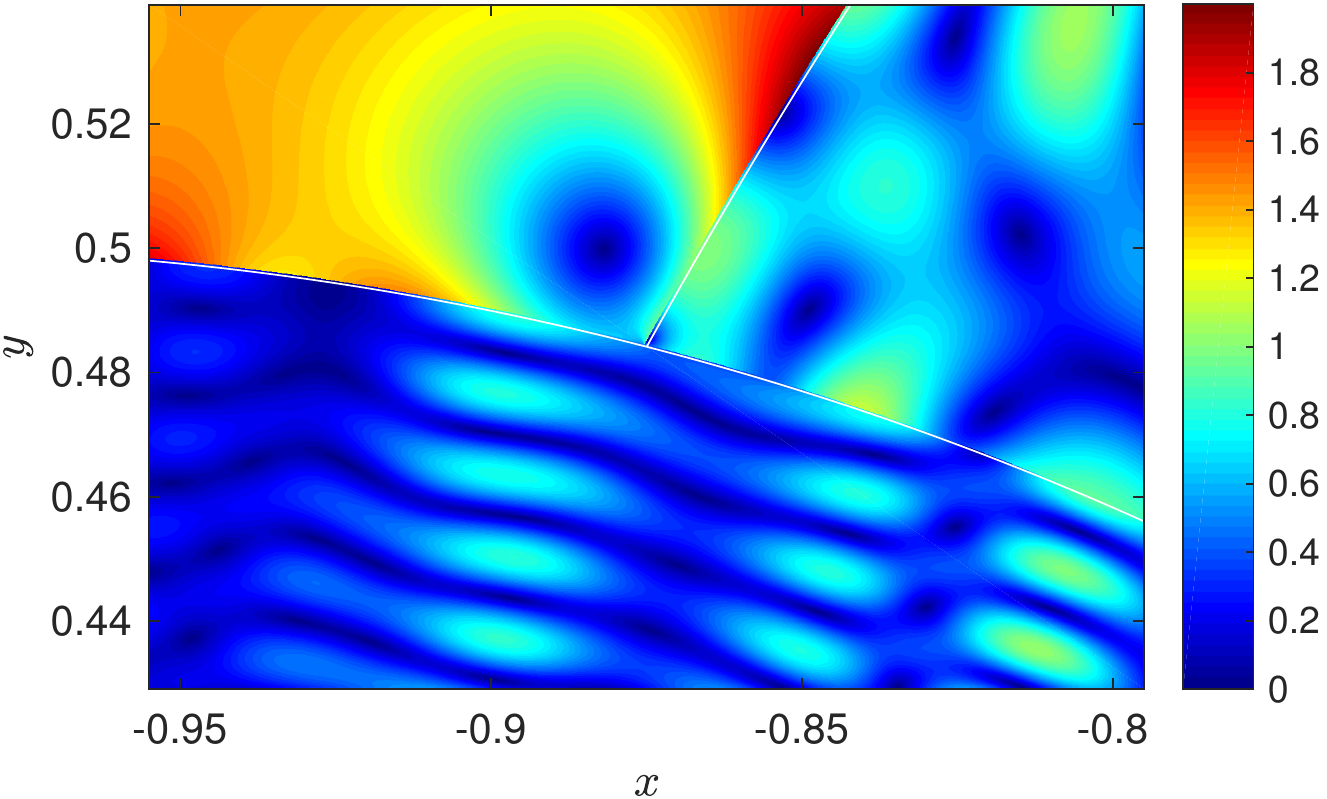}
\includegraphics[height=37mm]{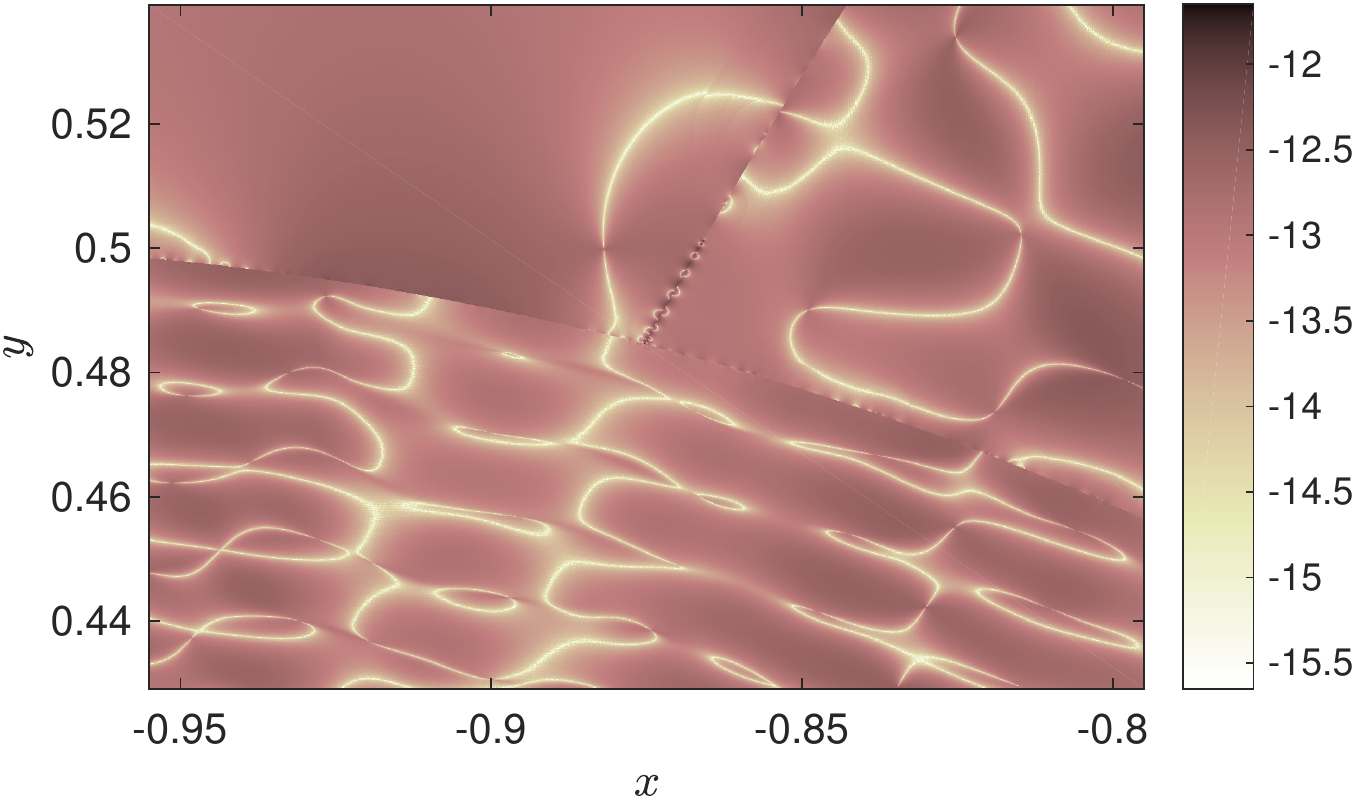}
\caption{\sf Same as the bottom row of Figure~\ref{fig:tricircUE}
  but the computational box is now ${\cal B}\approx\left\{-0.955\le
    x\le -0.795, 0.429\le y\le 0.539\right\}$.}
\label{fig:zoom}
\end{figure}

The results, shown in Figure~\ref{fig:zoom}, illustrate that the
extended representation~(\ref{eq:ErepMOD}), together with the other
features in our numerical scheme, allow for high achievable accuracy
for high-contrast problems also very close to subcurve triple
junctions.

\subsubsection{Convergence and asymptotic behavior}

\begin{figure}[t]
\centering
\includegraphics[height=50mm]{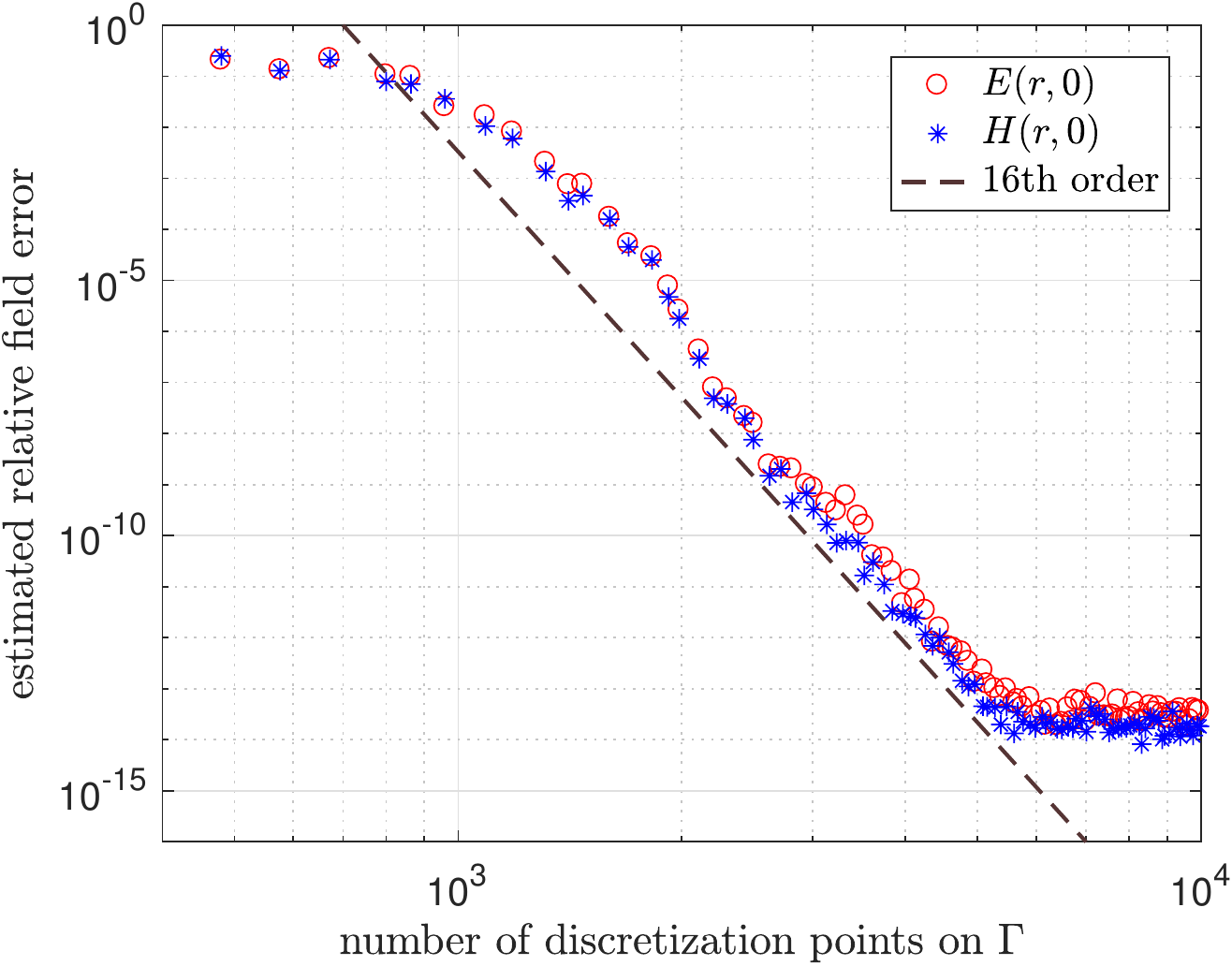}
\includegraphics[height=50mm]{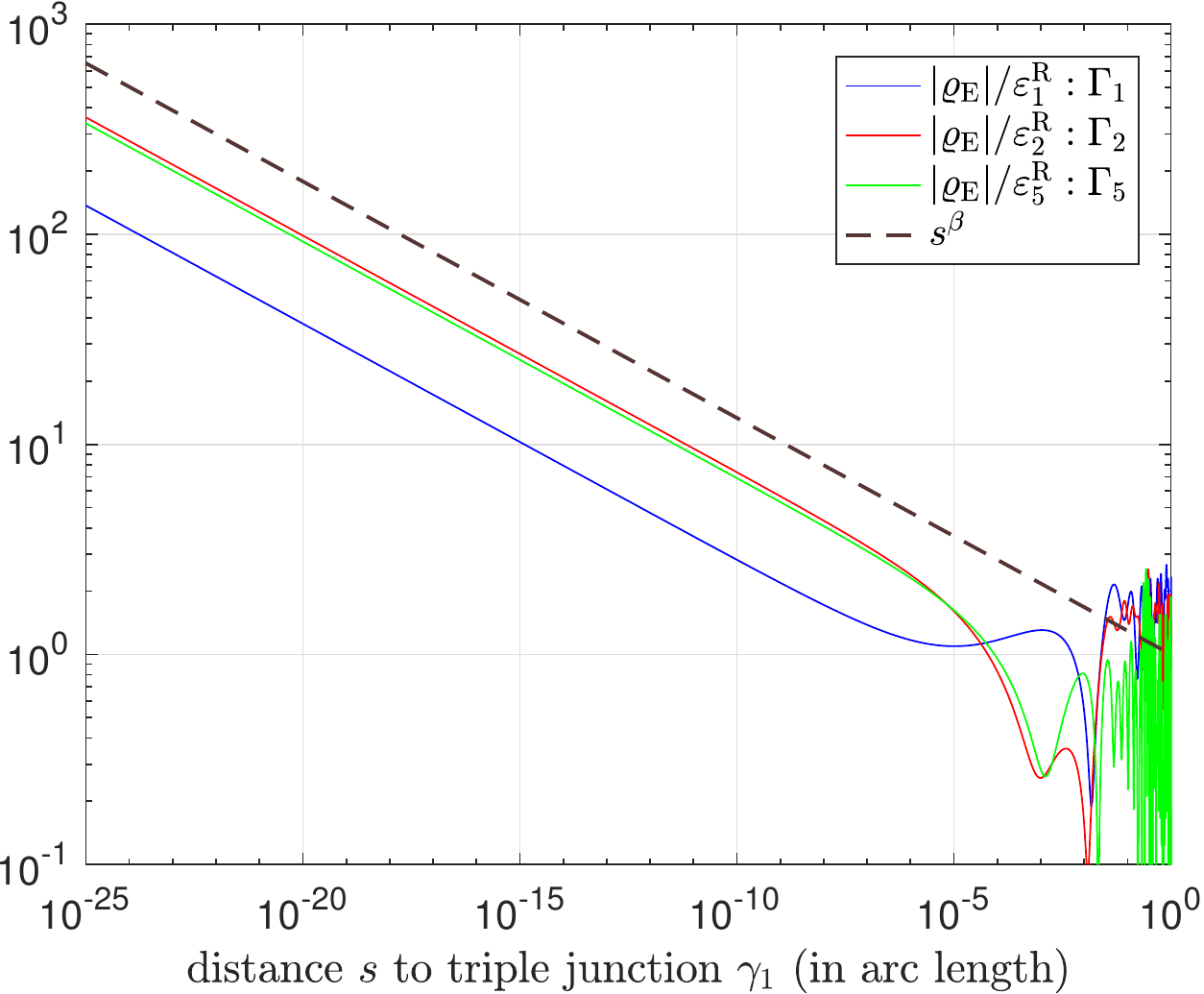}
\caption{\sf Left: convergence of $H_z(r,0)$ and $\lvert\myvec
  E(r,0)\rvert$, shown in Figure~\ref{fig:tricircUE}, as a function of
  the number of discretization points used on the coarse mesh on
  $\Gamma$. Right: behavior of $\lvert\rho_{\rm E}(r)\rvert$ close to
  the subcurve triple junction $\gamma_1$ in Figure~\ref{fig:zoom}.}
\label{fig:convasymp}
\end{figure}

Our discretization scheme uses composite $16$-point Gauss--Legendre
quad\-ra\-ture as underlying quadrature. If this was the only
quadrature used, the overall convergence of the scheme would be $32$nd
order. Since parts of the scheme rely on piecewise polynomial
interpolation, however, the overall convergence is $16$th order. This
is illustrated in the left image of Figure~\ref{fig:convasymp}, where
we show convergence for $\myvec H(r,0)$ and $\myvec E(r,0)$ with the
number of discretization points used on the coarse mesh on $\Gamma$.
The average estimated absolute field error is measured at $58,\!200$
points on a Cartesian grid in the box ${\cal B}=\left\{-1.6\le x\le
  1.6, -1.1\le y\le 1.1\right\}$ and normalized with the largest field
amplitude in ${\cal B}$. We use~(\ref{eq:rep1}) or~(\ref{eq:rep1sum})
for $\myvec H(r,0)$ and~(\ref{eq:Erepxy}) or~(\ref{eq:ErepxySUM}) for
$\myvec E(r,0)$. The reason for not using the extended
representation~(\ref{eq:ErepMOD}) of $\myvec E(r,0)$ is that it is
less memory efficient (in our present implementation) and that we need
heavily overresolved reference solutions for some data points.

The RCIP method lends itself very well to accurate and fully automated
asymptotic studies of surface densities close to singular boundary
points, see~\cite[Section~14]{Hels18}. As an example we compute
$\varrho_{\rm E}$ on $\Gamma_1$, $\Gamma_2$, and $\Gamma_5$ close to
the subcurve triple junction $\gamma_1$, which is zoomed-in in
Figure~\ref{fig:zoom}. The right image of Figure~\ref{fig:convasymp}
shows $\lvert\varrho_{\rm E}(s)\rvert$ as a function of the distance
$s$, in arclength, to $\gamma_1$. The leading asymptotic behavior is
$\lvert\rho_{\rm E}(s)\rvert\propto s^\beta$, with
$\beta=-0.1125730127414$, on all three subcurves.

\subsection{Two-region examples under plasmonic conditions}

We end this section by testing the new system~(\ref{eq:HKsys}). Recall
that the system~(\ref{eq:HKsys}), with $c$ as in~(\ref{eq:cchoice}),
has unique solutions on smooth $\Gamma$ when the
conditions~(\ref{eq:Klein}) hold. This includes {\it plasmonic
  conditions}. By this we mean that $k_1$ is real and positive and
that the ratio $\varepsilon_2/\varepsilon_1$ is purely negative or,
should no finite energy solution exist, arbitrarily close to and above
the negative real axis. Under plasmonic conditions, and when $\Gamma$
has corners, so called surface plasmon waves can propagate along
$\Gamma$. We will now revisit an example where this happens.

\subsubsection{Surface plasmon waves}

The example has $\Gamma$ given by~\cite[Eq.~(93)]{HelsKarl18}, shown
in Figure~\ref{fig:amoeba0}, and $k_0=18$, $\varepsilon_1=1$,
$\varepsilon_2=-1.1838$, $c=-{\rm i}$, and
$d=\left(\cos(\pi/4),\sin(\pi/4)\right)$. Results for $H_z^+(r,0)$ and
$\nabla H_z^+(r,0)$, where the plus-sign superscript indicates a limit
process for $\varepsilon_2/\varepsilon_1$, have been computed using an
abstract-density approach from~\cite{KleiMart88}
in~\cite[Figure~7]{HelsKarl18}. The main linear system in that
approach has~(\ref{eq:HKsysarrayadj2}) as its system matrix, so the
unique solvability is the same as for~(\ref{eq:HKsys}). The difference
between the two approaches lies in the representation formulas they
can use. The abstract-density approach is restricted to a local
representation of $U$~\cite[Eqs.~(22) and~(23)]{HelsKarl18} which
resembles~(\ref{eq:rep1}) and exhibits stronger singularities as
$\Omega_n\ni r\to r'\in{\cal C}_n$ than does the global
representation~(\ref{eq:rep1sum}), which can be used only together
with~(\ref{eq:HKsys}).

\begin{figure}[t]
\centering 
  \includegraphics[height=47mm]{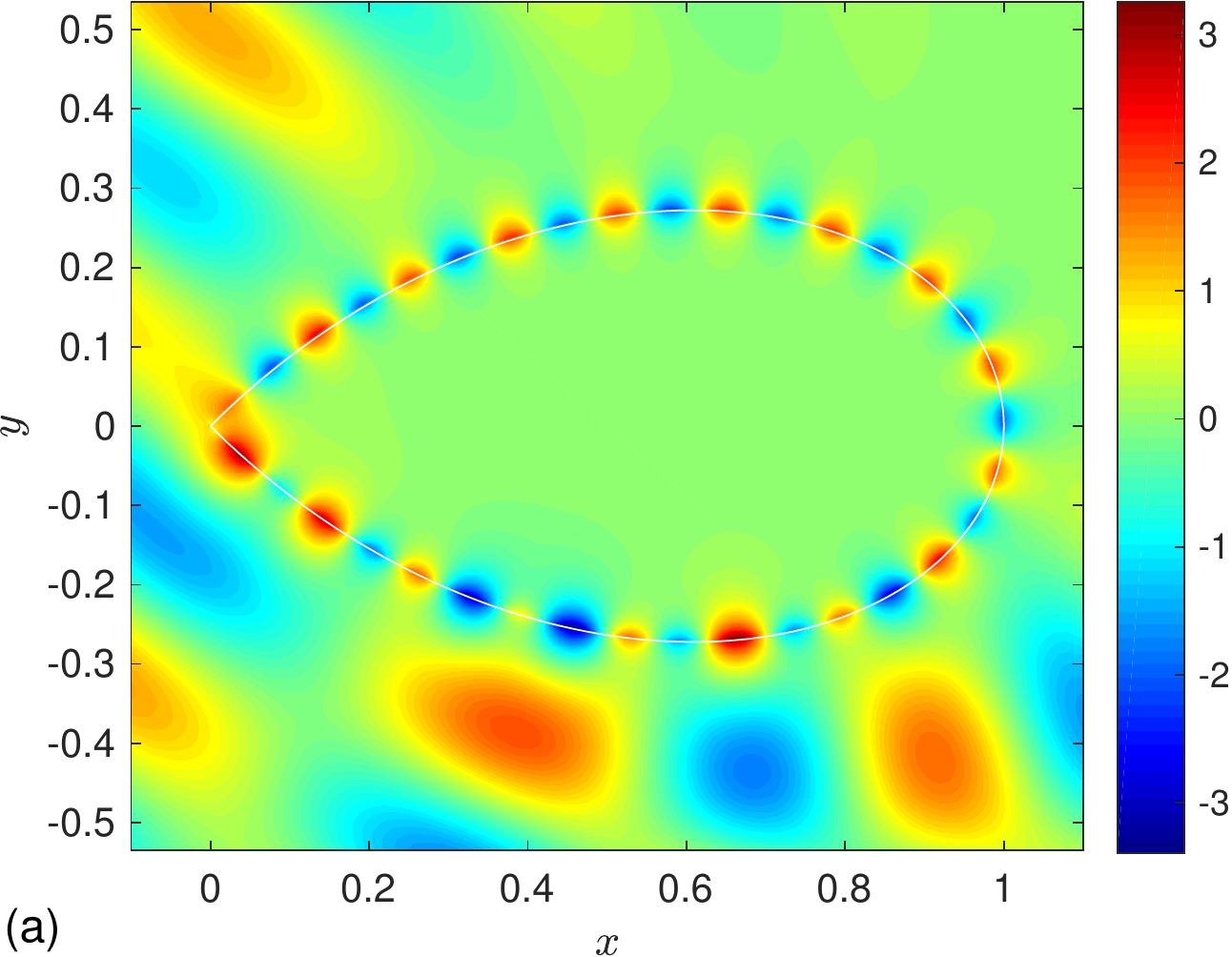}
  \includegraphics[height=47mm]{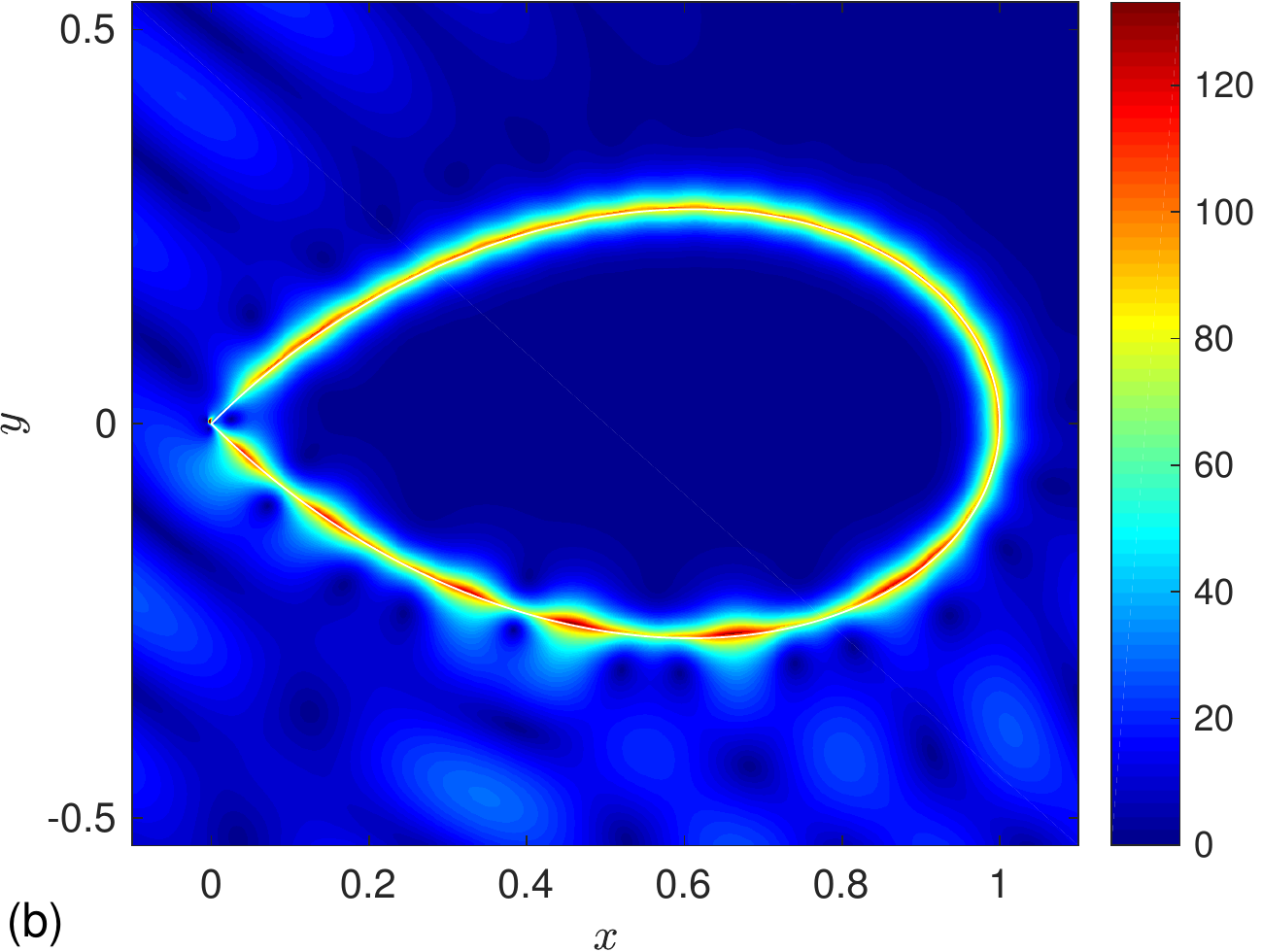}
  \includegraphics[height=46mm]{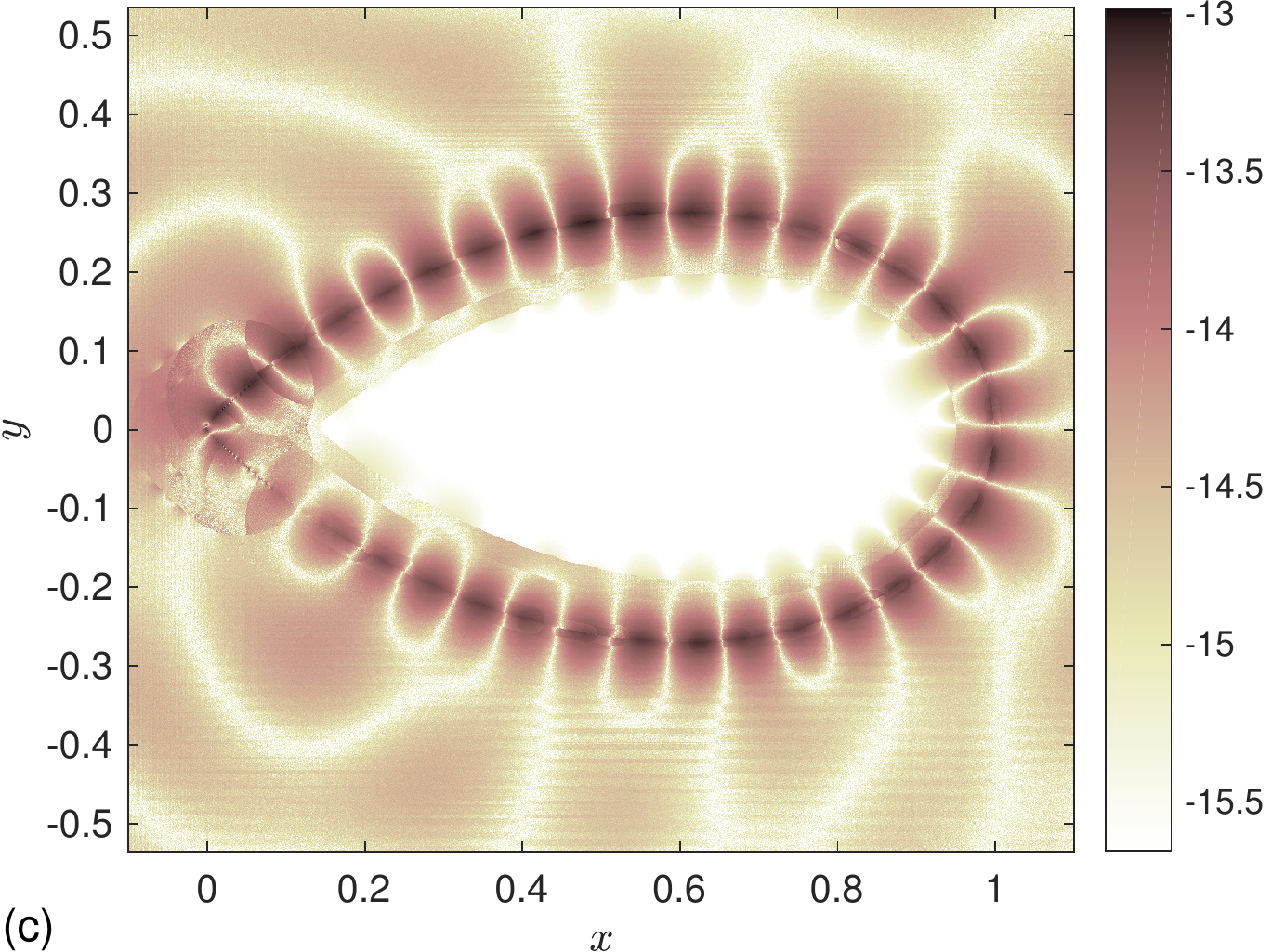}
  \includegraphics[height=46mm]{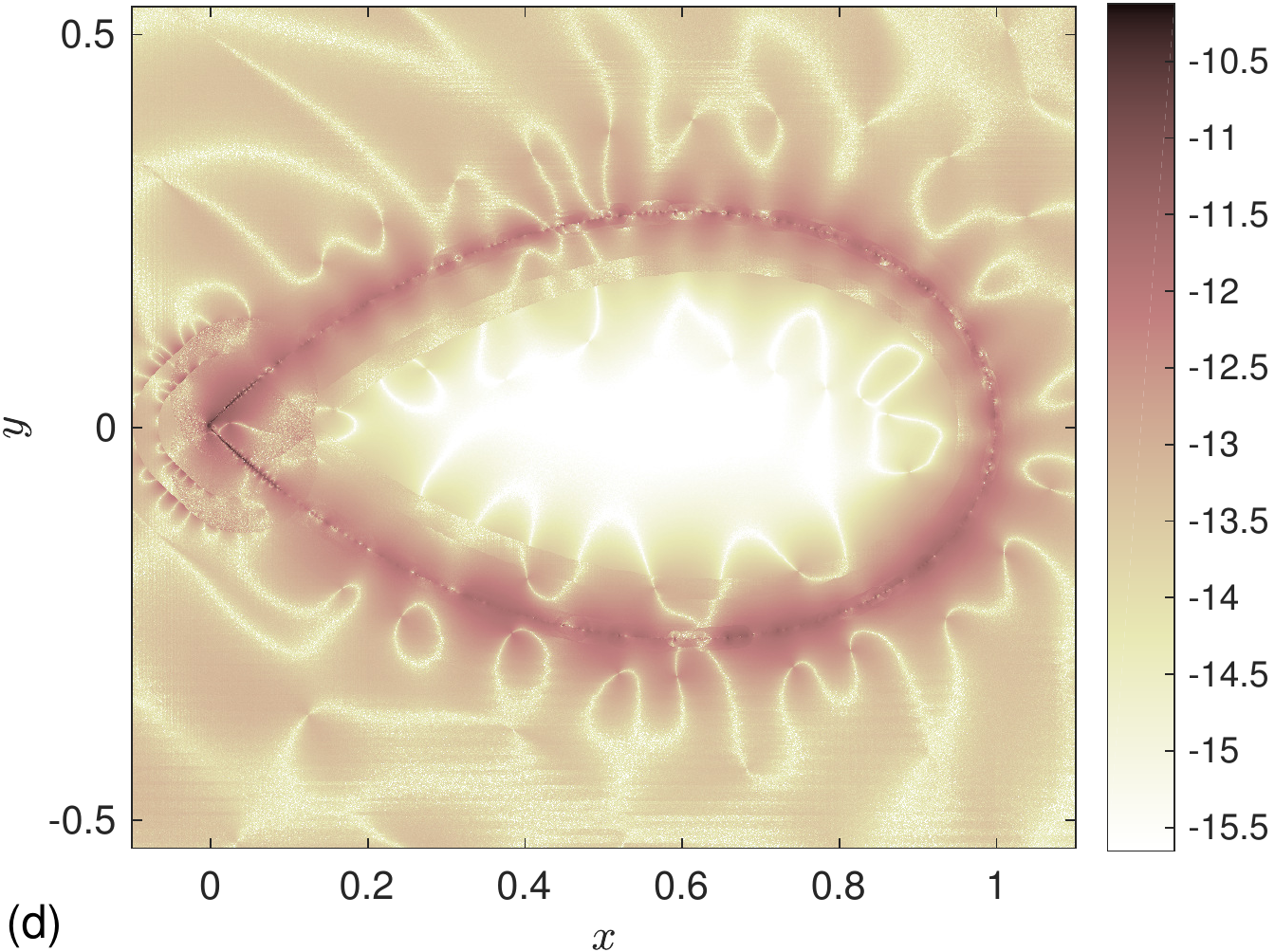}
\caption{\sf $H_z^+(r,0)$ and $\nabla H_z^+(r,0)$ 
  with $k_0=18$, $\varepsilon_1=1$, $\varepsilon_2=-1.1838$, and
  $d=\left(\cos(\pi/4),\sin(\pi/4)\right)$: (a) The field
  $H_z^+(r,0)$; (b) The (diverging) field $\lvert\nabla
  H_z^+(r,0)\rvert$ with colorbar range set to $[0,133]$; (c)
  $\log_{10}$ of estimated absolute error in $H_z^+(r,0)$; (d)
  $\log_{10}$ of estimated absolute error in $\lvert\nabla
  H_z^+(r,0)\rvert$.}
\label{fig:neg}
\end{figure}

Figure~\ref{fig:neg} shows result obtained with the
system~(\ref{eq:HKsys}) and the representations~(\ref{eq:rep1})
and~(\ref{eq:rep1sum}) and their gradients. There are $800$
discretization points on the coarse mesh on $\Gamma$ and $10^6$ field
points on a rectangular Cartesian grid in the box ${\cal B}=\left\{
  -0.1\le x\le 1.1, -0.54\le y\le 0.54\right\}$. Comparing
Figure~\ref{fig:neg} with~\cite[Figure~7]{HelsKarl18}, where the same
grid was used, one can conclude that the achievable accuracy for $r$
close to $\Gamma$ is improved with around half a digit in $H_z^+(r,0)$
and one and a half digits in $\nabla H_z^+(r,0)$.

\subsubsection{Unique solvability on the unit circle}
\label{sec:uniquex}

In this last example, the statements made about unique solvability in
Section~\ref{sec:unique} and Section~\ref{sec:compare} are illustrated
by four simple examples on the unit circle. The setup is the same as
in~\cite[Section~9.2]{HelsKarl18}, where the properties
of~(\ref{eq:HKsysarrayadj2}) were studied: $\varepsilon_1=1$,
$\varepsilon_2=-1.1838$, and the condition numbers of the systems
under study are monitored as the vacuum wavenumber varies in the
interval $k_0\in[0,10]$. A number of at least $20,\!000$ steps are
taken in the sweeps and the stepsize is adaptively refined when sharp
increases in the condition number are detected.

\begin{figure}[t]
\centering 
  \includegraphics[height=47mm]{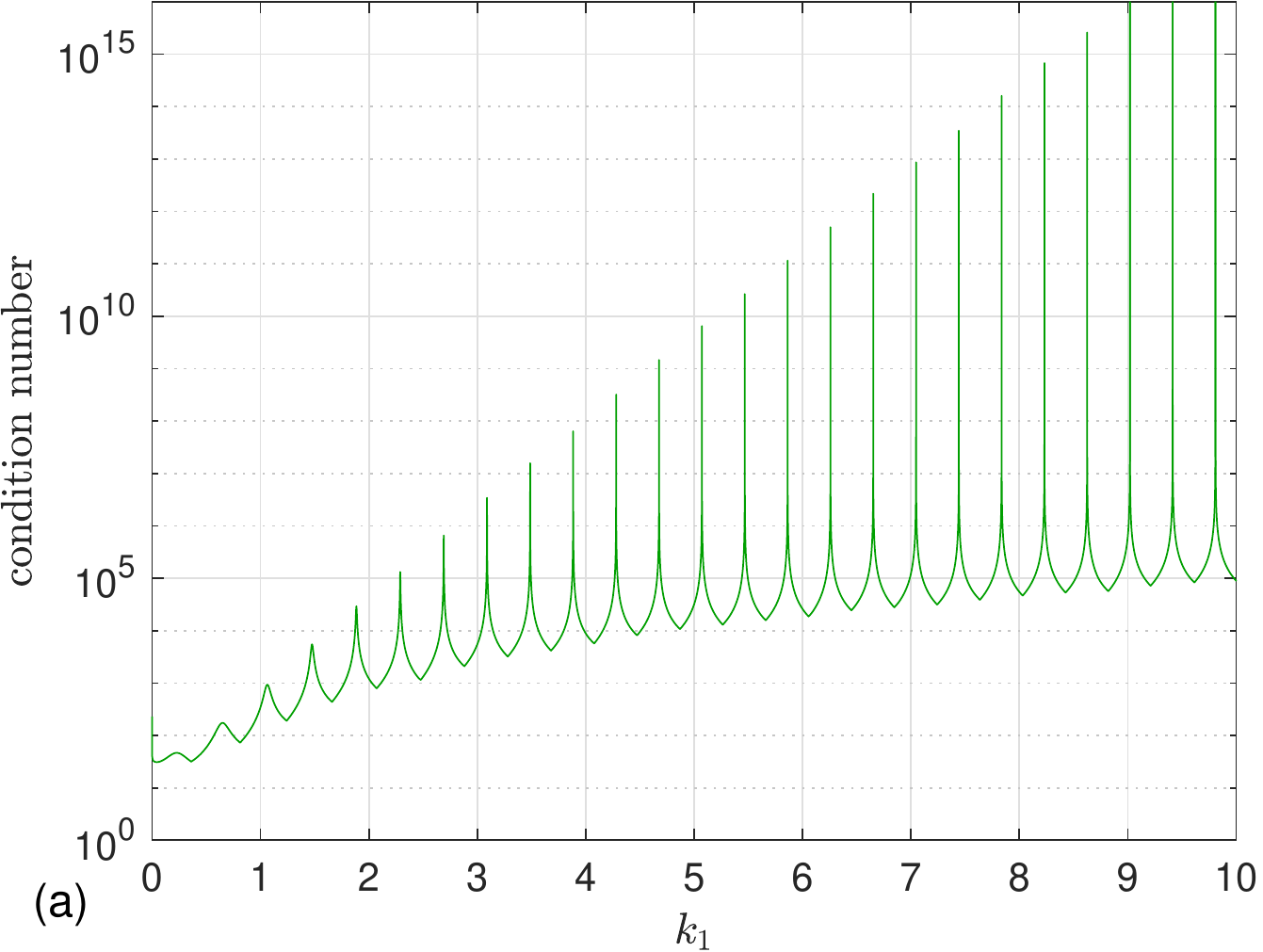}
  \includegraphics[height=47mm]{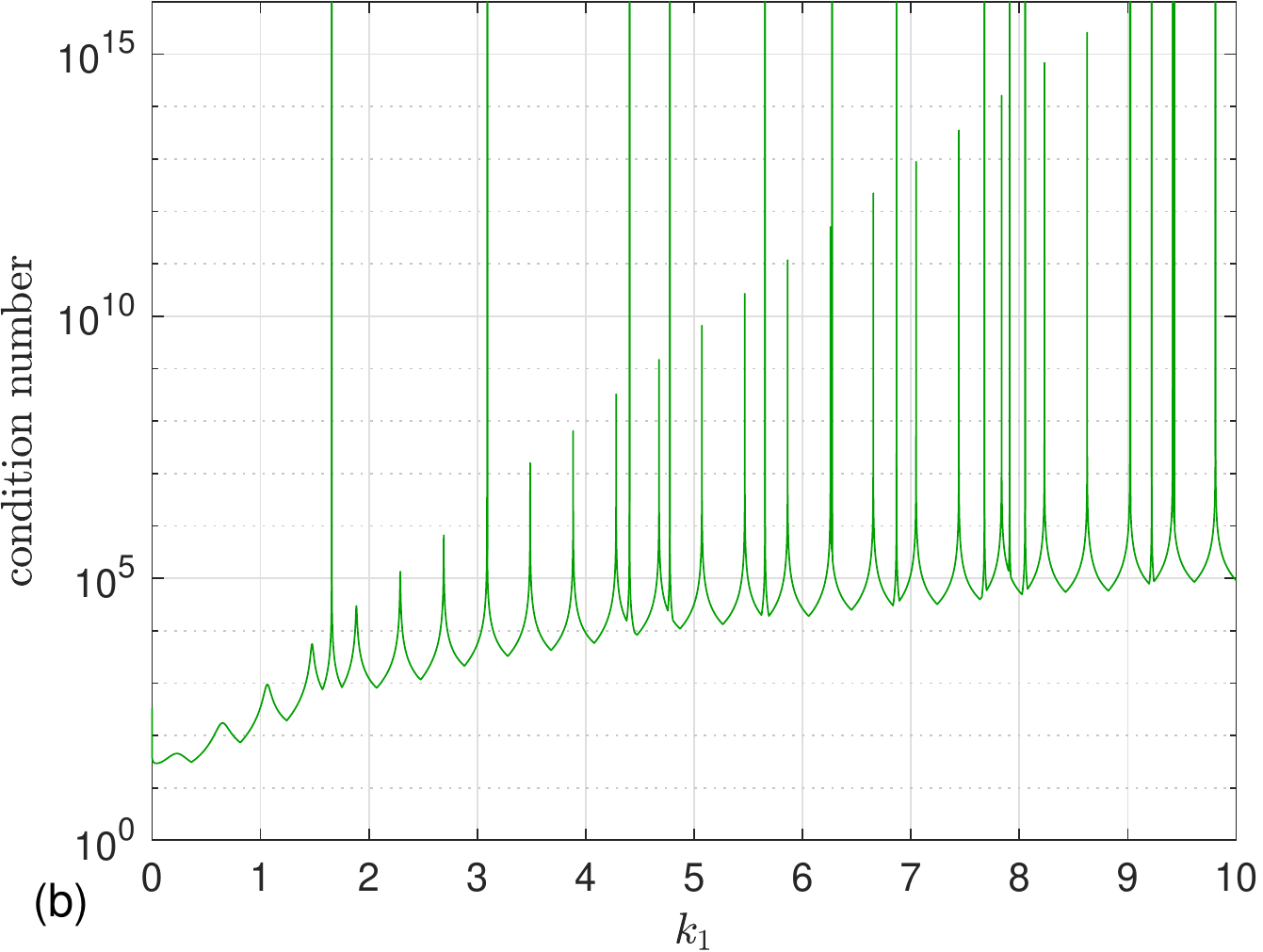}
  \includegraphics[height=47mm]{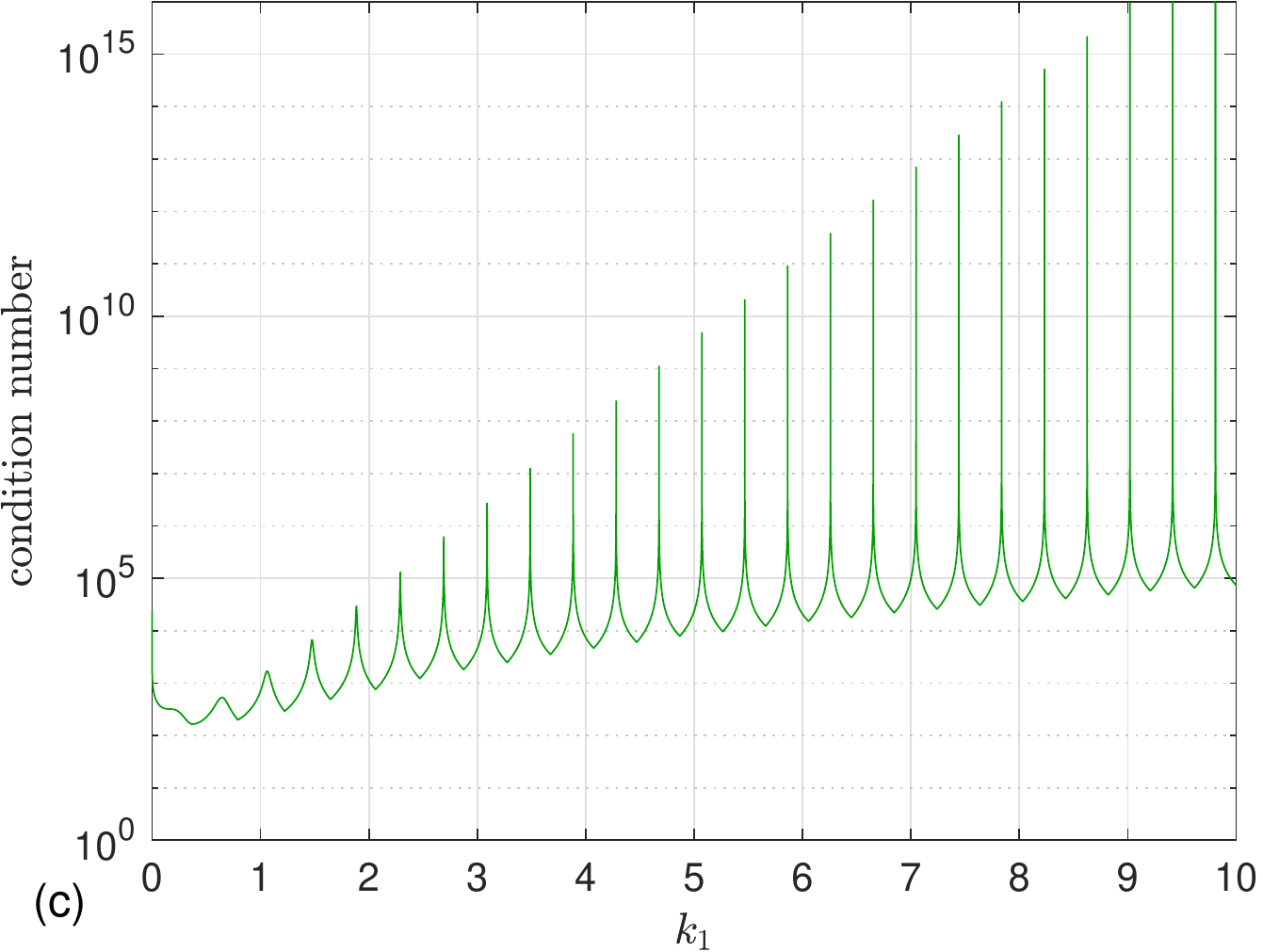}
  \includegraphics[height=47mm]{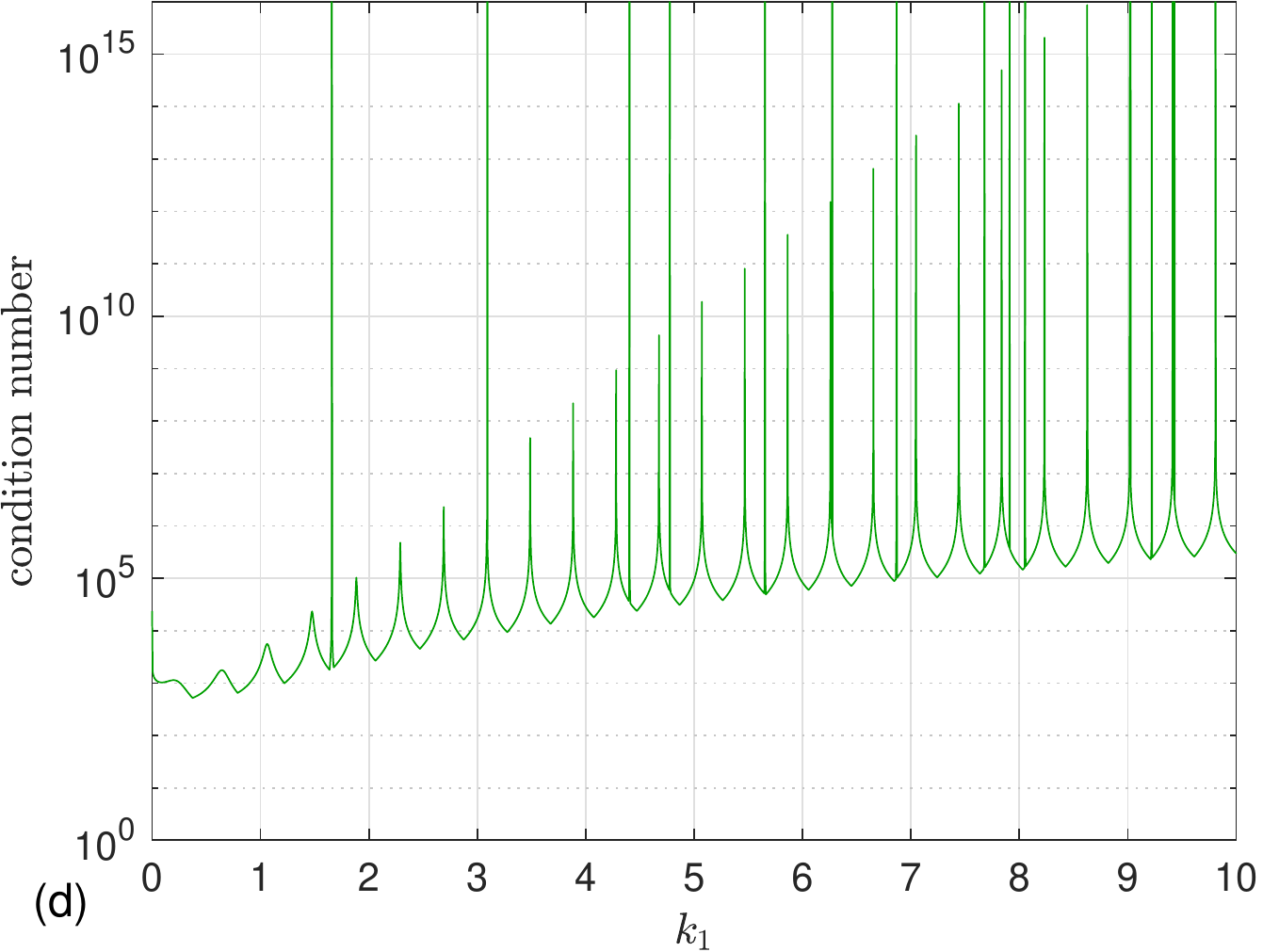}
\caption{\sf Condition numbers of system matrices on the unit 
  circle, $\varepsilon_2/\varepsilon_1=-1.1838$, and $k_1\in[0,10]$:
  (a) the system~(\ref{eq:HKsys}) with $c=-{\rm i}\,$; (b) the Müller
  system; (c) the ``$\myvec E$-system'' with $c=-{\rm i}\,$; (d) the
  original ``$\myvec E$-system''. The systems in (a,c) are free of
  false eigenwavenumbers while the systems in (b,d) exhibit twelve
  false eigenwavenumbers each.}
\label{fig:cond}
\end{figure}

We compare the system~(\ref{eq:HKsys}) with $c=-{\rm i}$, which is in
agreement with~(\ref{eq:cchoice}), to the Müller system adapted to the
problem of Section~\ref{sec:PDE}. The Müller system then corresponds
to~(\ref{eq:HKsys}) with $c=-1.1838$, according to
Section~\ref{sec:compare}. Figure~\ref{fig:cond}(a,b) shows that the
system~(\ref{eq:HKsys}) with $c=-{\rm i}$ is uniquely solvable in this
example while the Müller system has at least twelve wavenumbers
$k_1=k_0\in[0,10]$ where it is not. A number of $384$ discretization
points are used on $\Gamma$.

We also compare the two versions of the ``$\myvec
E$-system''~\cite[Eq.~(37)]{VicGreFer18}, adapted to the problem of
Section~\ref{sec:PDE} and discussed in Section~\ref{sec:compare}.
Figure~\ref{fig:cond}(c,d) shows that the ``$\myvec E$-system'' with
$c=-{\rm i}$ in the modified representation~(\ref{eq:vico23}) is
uniquely solvable in this example while the original ``$\myvec
E$-system'' with $c=-1.1838$ in~(\ref{eq:vico23}) has the same twelve
false eigenwavenumbers as the Müller system. A number of $768$
discretization points are used on $\Gamma$.

\section{Conclusions}
\label{sec:conclus}

Using integral equation-based numerical techniques, we can solve
planar multicomponent scattering problems for magnetic and electric
fields with uniformly high accuracy in the entire computational
domain. Almost all problems related to near-boundary field
evaluations, redundant contributions from distant sources, and
boundary subcurves meeting at triple junctions are gone. The success
is achieved through new integral representations of electromagnetic
fields in terms of physical surface densities, explicit kernel-split
product integration, and the RCIP method.

\section*{Acknowledgement}

\noindent
This work was supported by the Swedish Research Council under contract
621-2014-5159.

\begin{small}
\bibliography{helsbib}
\bibliographystyle{plain}
\end{small}

\end{document}